\documentclass[fleqn,usenatbib]{mnras}

\usepackage[T1]{fontenc}
\usepackage{ae,aecompl}
\usepackage{multirow}
\usepackage[normalem]{ulem}

\usepackage{graphicx}	
\usepackage{amsmath}	
\usepackage{amssymb}	
\usepackage[usenames, dvipsnames]{color}
\usepackage{orcidlink}

\usepackage{hyperref}

\newcommand{\msun}{\mbox{$\rm M_\odot$}}

\newcommand{\hagn}{\mbox{{\sc \small Horizon-AGN}}}

\newcommand{\nh}{\mbox{{\sc \small NewHorizon}}}

\newcommand{\orcid}[1]{\href{https://orcid.org/#1}{\textcolor[HTML]{A6CE39}{\aiOrcid}}}

\usepackage{soul}

\title[Black holes spins from NewHorizon]{Black hole spin evolution across cosmic time from the NewHorizon simulation}

\author[Beckmann, R. S.]{Beckmann, R. S. \orcidlink{0000-0002-2850-0192}$^{1,2}$\thanks{E-mail: ricarda.beckmann@roe.ac.uk },
Dubois, Y. \orcidlink{0000-0003-0225-6387}$^{3}$,
Volonteri, M. \orcidlink{0000-0002-3216-1322}$^{3}$,
Dong-Paez, C. A.\orcidlink{0000-0002-8590-4409}$^{3}$,
\newauthor
Peirani, S.\orcidlink{0000-0001-6902-2898}$^{3,4,5}$,
Piotrowska, J. M. \orcidlink{0000-0003-1661-2338}$^{6}$,
Martin, G. \orcidlink{0000-0003-2939-8668}$^{7}$,
Kraljic, K.  \orcidlink{0000-0001-6180-0245} $^{8}$,
\newauthor
Devriendt, J. \orcidlink{0000-0002-8140-0422}$^{9}$,
Pichon, C. \orcidlink{0000-0003-0695-6735}$^{3}$,
Yi, S. K. \orcidlink{0000-0002-4556-2619} $^{10}$
\\
$^{1}$ Institute of Astronomy and Kavli Institute for Cosmology, University of Cambridge, Madingley Road, Cambridge, CB3 0HA, UK\\
$^{2}$ Institute for Astronomy, University of Edinburgh, Royal Observatory, Edinburgh EH9 3HJ, UK\\
$^{3}$ Institut d’Astrophysique de Paris, CNRS, Sorbonne Université, UMR7095, 98bis bd Arago, 75014 Paris, France\\
$^{4}$ Université  et Observatoire de la Côte d’Azur, CNRS, Laboratoire Lagrange, Bd de l’Observatoire, 06304 Nice, France \\
$^{5}$ Department of Physics, School of Science, The University of Tokyo, 7-3-1 Hongo, Bunkyo-ku, Tokyo 113-0033, Japan\\
$^{6}$ Cahill Center for Astronomy and Astrophysics, California Institute of Technology, Pasadena, CA, USA \\
$^{7}$ School of Physics and Astronomy, University of Nottingham, University Park, Nottingham NG7 2RD, UK \\
$^{8}$ Observatoire Astronomique de Strasbourg, Universit\'e de Strasbourg, CNRS, UMR 7550, F-67000 Strasbourg, France \\
$^{9} $ Department of Physics, University of Oxford, Keble Road, Oxford, OX1 3RH, UK \\
$^{10}$ Department of Astronomy and Yonsei University Observatory, Yonsei University, Seoul 03722, Republic of Korea \\
}

\date{Accepted XXX. Received YYY; in original form ZZZ}

\pubyear{2024}

\begin{document}

\label{firstpage}
\pagerange{\pageref{firstpage}--\pageref{lastpage}}
\maketitle

\begin{abstract}
Astrophysical black holes (BHs) have two fundamental properties: mass and spin. While the mass-evolution of BHs has been extensively studied, much less work has been done on predicting the distribution of BH spins. In this paper we present the spin evolution for a sample of intermediate-mass and massive BHs from the \nh~ simulation, which evolved BH spin across cosmic time in a full cosmological context through gas accretion, BH-BH mergers and BH feedback including jet spindown. As BHs grow, their spin evolution alternates between being dominated by gas accretion and BH mergers. Massive BHs are generally highly spinning.Accounting for the spin energy extracted through the Blandford-Znajek mechanism increases the scatter in BH spins, especially in the mass range $10^{5-7} \rm \  M_\odot$, where BHs had previously been predicted to be almost universally maximally spinning. We find no evidence for spin-down through efficient chaotic accretion. As a result of their high spin values, massive BHs have an average radiative efficiency of $<\varepsilon_{\rm r}^{\rm thin}> \approx 0.19$. As BHs spend much of their time at low redshift with a radiatively inefficient thick disc, BHs in our sample remain hard to observe. Different observational methods probe different sub-populations of BHs, significantly influencing the observed distribution of spins. Generally, X-ray-based methods and higher luminosity cuts increase the average observed BH spin. When taking BH spin evolution into account, BHs inject on average between 3 times (in quasar mode) and 8 times (in radio mode) as much feedback energy into their host galaxy as previously assumed.
\end{abstract}




\section{Introduction}

Any astrophysical black hole (BH) can be described by two fundamental properties: its mass and its spin. Mass is the simpler of the two, accumulating through gas accretion and mergers but never decreasing \citep[if Hawking radiation is negligible,][]{Hawking1975}. Spin is more complex. As well as being a vector, with both a direction and magnitude evolving, BH spin can both increase and decrease throughout a BH's lifetime, through gas accretion \citep{Bardeen1970}, spin energy extracted by feedback \citep{Blandford1977} and BH-BH mergers \citep{rezzollaetal08,barausse_predicting_2009}.

The spin of a BH is tightly linked to how efficiently it converts accreted mass to radiation:
\begin{equation}
    L_{\rm BH} = \varepsilon_{\rm r}(a) \dot{M}_{\rm BH} c^2
    \label{eq:Lbol}
\end{equation}
where $L_{\rm BH}$ is the BH luminosity, $\dot{M}_{\rm BH}$ is the BH mass accretion rate, $c$ is the speed of light and $\varepsilon_{\rm r}(a)$ is the radiative efficiency of the accretion disc and $a$ is the spin magnitude. $\varepsilon_{\rm r}$ depends on spin as it is determined by the structure of the accretion disc close to the BH, which in turn depends on the BH spin magnitude and direction relative to the accretion disc spin and magnitude. How BH spin is changed during gas accretion depends on the current spin of the BH, the angular momentum of the accreted gas and the angle between the two \citep{BardeenPetterson1975,kingetal05,Lodato2006}. Accretion can drive spin to a maximum value of $a=0.998$ \citep{Thorne1974}, but there is some more recent theoretical work suggesting that this value could be exceeded for BHs accreting at super-Eddington rates \citep{Sadowski2011}. The spin of two BHs during the merger has important consequences for the velocity kick received by the remnant following a BH-BH merger \citep{lousto_statistical_2010,lousto_gravitational_2012}.

As the BH spin evolves continuously throughout a BH's lifetime, building a nuanced understanding of the long-term evolution of BHs, and their impact on their host galaxies, requires following BH mass and BH spin evolution over cosmological timescales. Early studies using semi-analytic models based on galaxy merger trees or analytical arguments showed that BH spin values depend strongly on the (an)isotropy of gas accreted by the BH throughout its lifetime \citep{Volonteri2005,Shapiro2005,King2008,BertiVolonteri2008,Fanidakis2011}. These early works concluded that randomly oriented gas accretion (chaotic accretion) leads to low spin values ($a=0.1-0.3$ but with significant scatter), while coherent accretion leads to maximally spinning BHs ($a > 0.9$). \citet{BertiVolonteri2008} concluded that gas accretion generally dominates over the contribution of BH-BH mergers, and that mergers drive spins $>0.7$. \citet{Volonteri2013} expanded on this work and showed that at high redshift, BHs tend to spin up rapidly due to efficient gas accretion. At low redshift, BH spins for massive BHs decrease again due to dry BH-BH mergers, while lower-mass BHs tend to be spun down by chaotic accretion. \citet{Dotti2013} used a parameter study to show that low-mass and high-mass BHs respond differently to anisotropy in the accreted gas: as low-mass BHs have short re-alignment timescales, they tend to be spun up maximally but show erratically changing BH spin direction when encountering chaotic accretion. More massive BHs do not significantly change their orientation during a given accretion event, and as a result their spin value reflects the coherence of the accreted gas: high spin for coherent accretion, low spin for chaotic accretion. Using semi-analytic models, \citet{Sesana2014} confirm that while BHs in the mass range $M_{\rm BH} = 10^6 - 10^7 \rm \ M_\odot$ are maximally spinning, average BH spin values decrease while scatter increases for higher BH mass. \citet{griffin_evolution_2019} confirm these trends with BH mass for both chaotic and coherent accretion models and \citet{IzquierdoVillalba2020} focus on the connection between spin and galaxy morphology, showing that BHs in elliptical galaxies have systematically lower spin than those in pseudo or classical bulges. \citet{Maio2013} demonstrated that stellar feedback insufficiently randomises gas flows in disc-like galaxies to spin down BHs: even in the presence of feedback, they report BH spins of $0.6-0.9$ regardless of initial BH spin.

Pioneering work using cosmological simulations was conducted by \citet{Dubois2014spin}. Using cosmological zoom simulations they showed that BHs spin up over cosmic time, but can have long periods of fixed spin magnitude at $z>3.5$ as their growth is regulated by the supernovae (SN) in their host galaxy. At lower redshift, BHs tend to be highly spinning ($a>0.7$) and spin back up efficiently after their spin is temporarily reduced during galaxy mergers. Using a larger sample of BHs and galaxies  \citet{Dubois2014} showed that generally, BH spin for BHs evolved over cosmological timescales remains high except for the most massive BHs in the local Universe ($z<0.5$) where average spin values drop by about 10 per cent. This is because most BHs are gas-accretion dominated, and BHs quickly spin up again following reductions in BH spin due to BH-BH and galaxy mergers.  This result was confirmed by \citet{Bustamante2019}, who also noted that self-consistently tracking BH spin evolution increases the scatter in the $M_{\rm star} - M_{\rm BH}$ relation, where $M_{\rm star}$ is the galaxy stellar mass. \citet{Beckmann2023a} showed that the distribution of BH spins depends at least in part on the evolution history of its host galaxy, with BHs in merger-free galaxies having on average higher spin values than those in merger-dominated galaxies. 

There have also recently been several efforts to study the evolution of BH spin using sub-grid accretion disc models that continuously model the evolution of the accretion disc, rather than assuming each accretion event is independent. \citet{Fiacconi2018} present a model for a sub-grid alpha disc evolving the BH spin, and find that for BHs with mass $M_{\rm BH} > 10^{8} M_\odot$, the angular momentum of the BH generally dominates over that of the disc, which makes spin flips more likely and accretion more episodic. For lighter BHs, the opposite is generally true and accretion is more steady. \citet{cenciBlackHoleSpin2020} used a similar model and show that the long-term evolution of the BH spin depends significantly on the radius at which the gas is assumed to circularise. \citet{koudmaniUnifiedAccretionDisc2024} present a model that seamlessly transitions from an alpha-disc at high accretion efficiencies to an ADIOS-like accretion flow at low efficiencies and find that the two-disc model significantly influences the final BH spin magnitude and direction. Finally, \citet{bollatiConnectionAGNRadiative2023} build on the model by \citet{cenciBlackHoleSpin2020} to study the impact of spin-driven feedback on BH evolution and conclude that in most situations it is comparatively small.

A BH's spin influences its ability to grow in mass in two ways: directly, as 
\begin{equation}
\dot{M}_{\rm growth} = (1 - \varepsilon_{\rm r}) \dot{M}_{\rm BH},
\label{eq:mass_growth}
\end{equation} 
i.e. any mass not converted to radiation directly contributes to the BH mass growth rate $\dot{M}_{\rm growth}$, and indirectly, in the sense that higher radiative efficiencies imply higher amounts of feedback energy that lead to stronger self-regulation of the BH. \citet{zubovas_slow_2019} showed that both effects combine to allow slowly spinning BHs to grow up to 20 per cent more than highly spinning BHs in the same galaxy, which could have important consequences for the assembly of the first quasars \citep{zubovas_high-redshift_2021,pacucci_search_2021}. Super-Eddingon growth could play a dual role, delivering large amounts of mass growth while spinning down the BH through the jets expected to be present for the thick accretion discs that power super-Eddington growth. However, in
 \citet{massonneau_how_2023} we showed that such Super-Eddington growth can delay but ultimately not prevent the spinup of the BH. As a result, self-regulation remains too efficient and long-term BH growth remains stifled even for an initially non-spinning BH. The length of the delay depends strongly on the adopted jet feedback efficiency \citep{massonneau_how_2023b}.

Observationally, BH spin is much more difficult to measure than BH mass \citep[see][for a recent review]{reynolds_observational_2020}. In recent years, observations of massive BH spins have come predominantly from a variety of methods. The most direct probe is X-ray reflection spectroscopy \citep[e.g.][]{Brenneman2011,reynolds13,mallick_high-density_2022,sisk-reynes_evidence_2022}, but this requires long exposure times. For larger samples of active galacit nuclei (AGN), BH spin has been estimated more coarsely from the jet luminosity \citep{daly_estimates_2011,Daly2019}, the radiative efficiency \citep{Trakhtenbrot2014} and continuum fitting to the hard X-ray spectrum \citep{you_constraints_2016}. While this body of work gives an exciting glimpse into the distribution of BH spins in the Universe, the overall distribution of BH spins remains poorly understood as observational sample sizes are small, current observational methods  are sensitive to assumptions on the structure of the accretion disc \citep{riaz_modeling_2020} and remain biased towards high spin values \citep{vasudevan_selection_2016}.

In this paper, we present a study of the long-term spin evolution of intermediate-mass and massive BHs in the \nh~ simulation. The paper is structured as follows: After introducing the simulation in Sec. \ref{sec:simulation}, we present results on the spin-evolution of BHs in Sec. \ref{sec:spin}, with a specific focus on the impact of BH-BH mergers and gas accretion in Sec. \ref{sec:mergers} and \ref{sec:gas_accretion} respectively. The impact of BH spin on the luminosity function is discussed in Sec. \ref{sec:radiative_efficiency}, and the potential observability of BH spin in Sec. \ref{sec:observability}. Conclusions can be found in Sec. \ref{sec:conclusions}.

\section{Simulation}
\label{sec:simulation}

\nh~is a high-resolution resimulation of an average sub-volume of the \hagn~simulation \citep{Dubois2014}. \nh~has been presented in detail in \citet{Dubois2021}.

\nh~consists of a high-resolution region with a radius of 10 comoving Mpc, with a DM mass resolution of $ 1.2 \times 10^6 \msun$ which is embedded  within the 142 comoving Mpc box of \hagn. It uses a $\Lambda$CDM cosmology consistent with WMAP-7 data \citep{komatsu2011} with a total matter density $\Omega_{\rm m}=0.272$, a dark energy density $\Omega_{\Lambda}=0.728$, a baryon density $ \Omega_{\rm b}=0.045$ and  a Hubble constant of $H_{0}=70.4 \rm \ km\,s^{-1}\, Mpc^{-1}$. The amplitude of the matter power spectrum and power-law index of the primordial power spectrum are $\sigma_8=0.81$ and $n_{\rm s}=0.967$ respectively. 

\nh~was performed with {\sc ramses} \citep{Teyssier2002}, using a second-order unsplit Godunov scheme for solving the Euler equations, and an HLLC Riemann solver with a MinMod Total Variation Diminishing  scheme  to  reconstruct  interpolated  variables. Refinement follows a quasi-Lagrangian scheme where a cell is refined if its mass exceeds 8 times the initial mass distribution, up to a maximum resolution of $34 \rm \ comoving \ pc$. The minimum cell size is kept approximately constant throughout by adding an extra level of refinement at expansion factor $a_{\rm exp}=0.1,0.2,0.4$ and $0.8$. We supplement this quasi-Lagrangian scheme with a super-Lagrangian refinement criterion in cells with a gas number density larger than $5\, \rm H\, cm^{-3}$, which enforces refinement of cells whose size is smaller than one Jeans length.

Gas follows an ideal monoatomic equation of state with an adiabatic index of $\gamma_{ad}=5/3$, and cooling is modelled assuming equilibrium chemistry using cooling curves from \citet{Sutherland1993} down to $10^4\, \rm K$.  Heating from a uniform UV background takes place after redshift $z_{\rm reion} = 10$ following \citet{Haardt1996}. Stars are formed in cells with a gas number density above $n_0=10\, \rm H\, cm^{-3}$, following a Schmidt relation \citep{kimmetal17,trebitschetal17,trebitschetal20} and using a star formation efficiency that depends on the properties of the interstellar medium such as the Mach number and virial parametres. This leads to a stellar mass resolution of $ 1.3\times 10^4 \ \msun$). Stars are assumed to have a Chabrier \citep{Chabrier2005} initial mass function with cutoffs at 0.1 and 150 \msun, and stellar feedback separately tracks the momentum and energy-conserving phase of the explosion following \citet{Kimm2015}.

\subsection{Black hole formation, dynamics and mergers}
\label{sec:BHform_dyn_merger}

BHs are formed in cells whose gas and stellar density exceeds the threshold for star formation if the cell also has a stellar velocity dispersion of more than 20 $\rm km s^{-1}$ and is located at least 50 kpc from any existing BH. BHs form with a mass of $10^4 \ \msun$. To avoid spurious motions of BHs due to finite force resolution effects, we include an explicit drag force of the gas onto the BH, following the analytic description from \citet{Ostriker1999}. BHs are merged when their relative velocity is smaller than the escape velocity of the binary, and when they approach closer than $4\Delta x$ ($\sim 150$~pc). The resulting distribution of BHs mergers in \nh~ is analysed in \citet{Volonteri2020}. We do not model post-merger velocity kicks here.

\subsection{Black hole accretion model} 
\label{sec:BHacc}

To model the evolution of BH spin through gas accretion and feedback we use a two-disc model, tied to the two-mode feedback model: BHs accrete gas following un-boosted Bondi-Hoyle-Lyttleton accretion $\dot{M}_{\rm BHL}$, based on local mass-weighted, kernel-weighted gas quantities. During each accretion event, the angular momentum of the accreted gas is measured, using the same kernel-weighting as for the accretion. Accretion is capped at the Eddington rate $\dot{M}_{\rm Edd}$, which is computed using the spin-dependent radiative efficiency $\varepsilon^{\rm thin}_{\rm r}$ assuming a thin accretion disc following \citet{Bardeen1970} such that the accretion rate onto the BH is $\dot{M}_{\rm BH} = \min(\dot{M}_{\rm BHL},\dot{M}_{\rm Edd})$. For efficiently growing BHs (Eddington ratio $f_{\rm Edd} = \frac{\dot{M}_{\rm BHL}}{\dot{M}_{\rm Edd}} > 0.01$), the effective radiative efficiency $\varepsilon_{\rm r}$ used to determine the distribution of accreted mass between BH mass growth and feedback (see Eq. \ref{eq:mass_growth}) and the bolometric luminosity (see Eq. \ref{eq:Lbol}), is taken to be that of the thin disc: $\varepsilon_{\rm r} = \varepsilon^{\rm thin}_{\rm r}$, following \cite{Shakura1973}.

To model the impact of BHs transitioning to a thick accretion disc for low Eddington ratios, we follow \citet{benson&babul09} and attenuate $\varepsilon^{\rm thin}_{\rm r}$ by a factor $f_{\rm att}=f_{\rm Edd}/0.01$ for an effective radiative efficiency $\varepsilon_{\rm r} = f_{\rm att} \varepsilon^{\rm thin}_{\rm r}$ for BHs with an Eddington ratio $f_{\rm Edd} \leq 0.01$. During each accretion event, we measure the total mass accreted onto the BH, $M_{\rm acc,BH}$ as well as the angular momentum unit vector of the accreted gas $\mathbf{j}_{\rm gas}$.

\subsection{BH spin evolution model}
\label{sec:BHspin}
In \nh, BH spin is modelled on-the-fly and updated according to gas accretion, BH-BH mergers and BH spin-down during feedback. All BHs are seeded with zero spin, i.e. spin parameter $a = 0$. During each accretion event, we update the spin of the BH according to the angular momentum of the accreted gas, following the model first presented in \citet{Dubois2014spin} and updated in \citet{Dubois2021}, which we summarise here.

To update the BH spin during each accretion event, we first assume that the angular momentum direction of unresolved accretion disc $\mathbf{j}_{\rm D}$ is the same as that of the inflowing gas measured on simulation scales, i.e.  $\mathbf{j}_{\rm D}= \mathbf{j}_{\rm gas}$. We then measure the angle $\theta$ between $\mathbf{j}_{\rm D}$ and the current BH spin  unit vector $\mathbf{j}_{\rm BH}$. As we cannot resolve the structure of the accretion disc and its evolution explicitly, we iteratively update the spin of the BH as follows:

First we compute the warp radius of the disc $R_{\rm warp}$ and the mass of the disc within the warp radius, $M_{\rm d} = M_{\rm d} (R_{\rm warp})$. Then we check the radius at which the disc fragments due to self-gravity $R_{\rm sg}$. If $R_{\rm sg} < R_{\rm warp}$ we limit the disc mass to $M_{\rm d} = M_{\rm d} (R_{\rm sg})$. Finally we ensure mass conservation with the accreted mass by setting $M_{\rm d} = min(M_{\rm d}, M_{\rm acc, BH, remaining})$ where $M_{\rm acc, BH, remaining}$ is the total remaining mass to be accreted onto the BH during the current accretion event.

We then compute the disc angular momentum magnitude by integrating up to the warp radius, i.e. $J_{\rm d} \sim M_{\rm d} M_{\rm BH}^{1/2} R_{\rm warp}^{1/2}$ and the BH angular momentum magnitude $J_{\rm BH} = a M_{\rm BH}^{3/2} R_{\rm BH}^{1/2}$ where $M_{\rm BH}$ is the BH mass and $R_{\rm BH}^{1/2}$ is the BH's Schwarzschild radius. This allows us to evaluate whether the following criteria from  \cite{kingetal05} are fulfilled, in which case the disc is assumed to anti-align with the BH: $J_{\rm d} < 2 J_{\rm BH}$ and $\cos(\theta) < - \frac {J_{\rm d}}{2 J_{\rm BH}}$.

Finally we update the BH spin direction by aligning it with the total angular momentum of BH and disc, $\mathbf{J}_{\rm tot} = J_{\rm BH} + J_{\rm D} \mathbf{j}_{\rm D}$. The BH spin magnitude is updated using $M^{n+1}_{\rm BH} = M^{n}_{\rm BH} + M_{\rm d}$  according to one of two models: for a BH with an Eddington ratio of $f_{\rm Edd} > 0.01$, we assume a thin disc accretion model following \citet{Bardeen1970}, in which BHs are generally spun up (for $\mathbf{j}_{\rm D}$ aligned with $\mathbf{j}_{\rm BH})$. For BHs with $f_{\rm Edd} \leq 0.01$ we assume a thick disc solution, with spin-up rates and $\eta = \eta_{\rm radio} = \varepsilon_{\rm MCAD}$, where $\varepsilon_{\rm MCAD} $ takes the fourth-order polynomial form presented in \citet{Dubois2021} based on the results of the simulated magnetically chocked accretion discs (MCAD) from \citet{McKinney2012} who predict that BHs are generally spun down due to the rotational energy extracted to power their jets.

We iterate this sub-grid model until a total mass of $M_{\rm acc,BH}$ has been accreted onto the BH, i.e. until $M_{\rm acc,BH,remaining} = M_{\rm, acc, BH} - \sum_n M_{d,n}$ = 0 and the accretion event is completed.  Each accretion event is treated as the formation of a new accretion disc i.e. $M_{\rm acc, BH}$ and $\mathbf{j}_{\rm D}$ are independent between accretion events. This is unlike sub-grid accretion disc particle models, such as the ones used in \citet{Fiacconi2018, cenciBlackHoleSpin2020} and \citet{koudmaniUnifiedAccretionDisc2024} where the evolution of the disc is tracked continuously from one timestep to the next. All formulas used for $R_{\rm warp}$, $R_{\rm sg}$, $M_{\rm D}$ and the change in spin magnitude can be found in \cite{Dubois2014spin}.

During BH-BH mergers, the spin of the remnant is calculated according to the spin of the initial BHs, and the angular momentum of the binary, following \citet{rezzollaetal08}, but the orbital angular momentum of the binary is taken as random.

\subsection{BH feedback}
\label{sec:BHfeed}
After each accretion event, BH feedback energy is released in the form $\dot{E}_{\rm AGN} = \eta \dot{M}_{\rm BH} c^2$, where $\eta $ is the efficiency with which accreted energy couples to the interstellar medium (ISM). The feedback proceeds according to the same two-disc model also used for accretion (Sec. \ref{sec:BHacc}) and BH spin evolution (Sec \ref{sec:BHspin}):

For a BH with an Eddington ratio of $f_{\rm Edd} > 0.01$, we assume a thin disc accretion model and following \citet{Bardeen1970} and release BH feedback energy in the form of an isotropic injection of thermal energy with a feedback efficiency $\eta = \eta_{\rm quasar}=\varepsilon_{\rm f}\varepsilon_{\rm r}^{\rm thin}$, with $\varepsilon_{\rm f}=0.15$ (calibrated on the BH-galaxy mass relation). For BHs with $f_{\rm Edd} \leq 0.01$ we assume a thick disc solution, with spin-up rates and $\eta = \eta_{\rm radio} = \varepsilon_{\rm MCAD}$, where $\varepsilon_{\rm MCAD} $ takes the fourth-order polynomial form presented in \citet{Dubois2021} based on the results of the simulated magnetically chocked accretion discs (MCAD) from \citet{McKinney2012}. In this mode, feedback energy is injected in the form of bipolar kinetic jets with a spin-dependent efficiency $\eta$ that follows the results of~\citet{McKinney2012}.

\section{Black hole and galaxy catalogue}
The catalogue of BHs analysed here consists of all BHs in \nh~that are associated with host galaxies, which in turn have been identified with uncontaminated host halos, at redshift $z=0.25$. Uncontaminated halos are those with high-resolution DM particles only within their virial radius.

The catalogue of DM halos consists of all DM halos identified by the structure-finding algorithm HOP that only contain DM particles from within the zoom region, i.e. that are at the maximum DM resolution. Galaxies are identified from the star particles of the simulation using the HOP algorithm. The galaxy catalogue consists of all main or central galaxies located within the central 0.1 virial radii of a DM halo which have a stellar mass above $10^ 8 \rm \ M_\odot$. 

To identify a galaxy's main BH, we identify the most massive BH to be contained within 2 effective radii of the galaxy's centre, cycling over galaxies from the most to the least massive. This BH is flagged as the galaxy's main BH and removed from the sample of unallocated BHs. We then loop over all galaxies from most to least massive for a second time, identifying all remaining unallocated BHs contained within 2 effective radii of the galaxy as secondary BHs. A galaxy can contain multiple BHs, but a BH can only be associated with a single galaxy. Therefore the full sample of BHs discussed here contains all BHs associated with host galaxies. We exclude BHs from the sample that are contained in contaminated galaxies and halos and ``wandering'' BHs that are far from any galaxy. While this approach generally gives very satisfying matching between BHs and galaxies, it can artificially exclude BHs if their host galaxies have not been correctly identified by HOP. In the paper presented here, one of the most massive BHs, BH1049 was excluded by the automatic matching algorithm. After manually confirming that it is hosted in an uncontaminated galaxy, we have added BH1049 to our catalogue of BHs. The catalogue of BHs analysed here is the same as in \citet{Beckmann2023} except for the additional BH 1049 which was omitted in the previous paper.

\section{Results}

\subsection{The long-term spin evolution of black holes}
\label{sec:spin}

\begin{figure}
    \centering
    \includegraphics[width=\columnwidth]{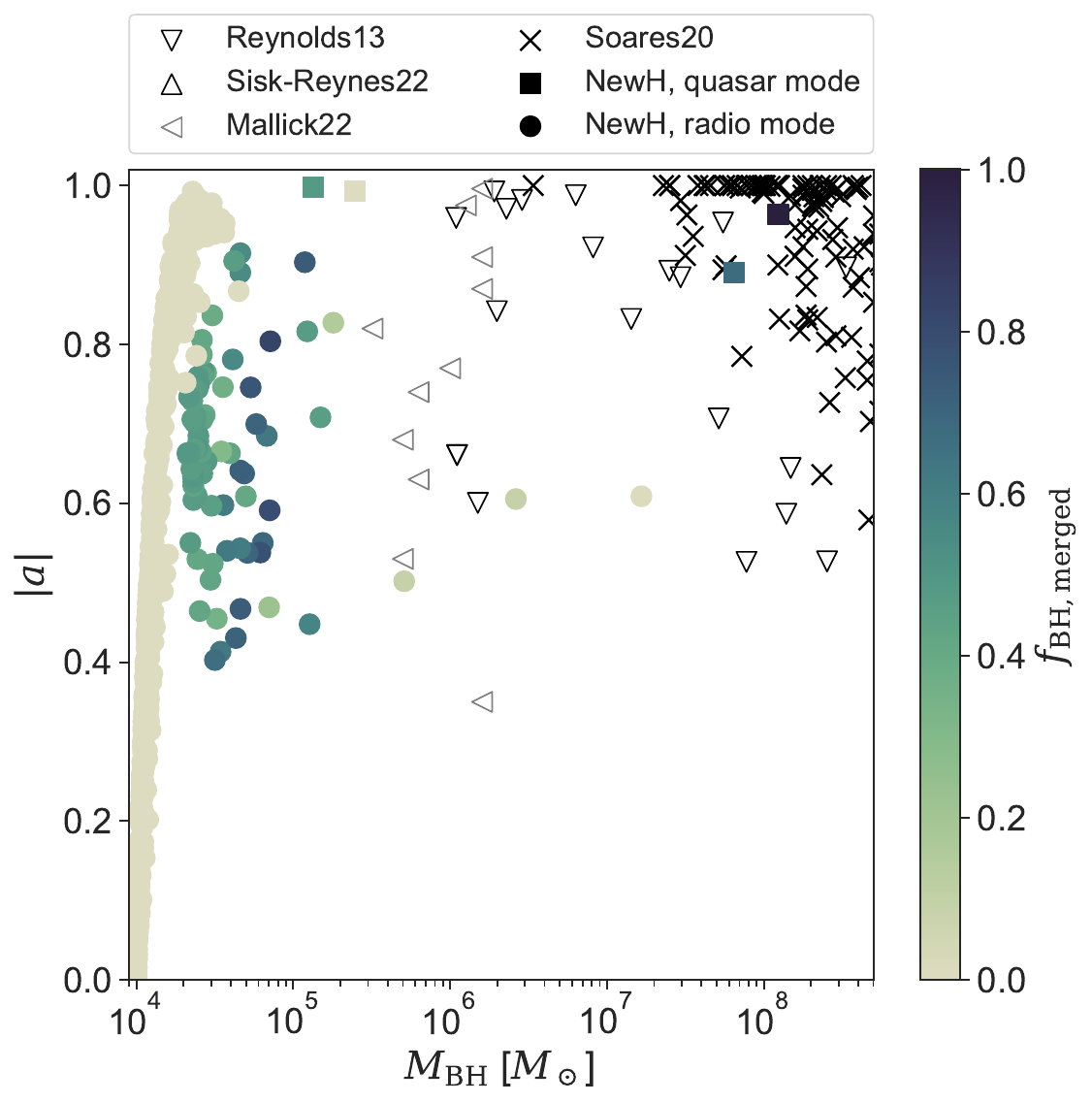}
    \caption{Distribution of BH spins versus BH mass for all $z=0.25$, colour-coded by the fraction of that BH's mass that has been acquired via mergers, $f_{\rm BH, merged}$. BH spin observations from \citet{reynolds13}, \citet{sisk-reynes_evidence_2022}, \citet{mallick_high-density_2022} and \citet{soares&nemmen20} are shown with open markers. Square markers denote BHs in quasar (thin disc) mode, while round markers denote BHs in radio (jetted, thick disc) mode in the simulations. All observations are for BHs in quasar mode.}
    \label{fig:spin_merger_mass}
\end{figure}

 Fig.~\ref{fig:spin_merger_mass} shows the distribution of absolute BH spin magnitudes $|a|$ 
 for all BHs in \nh~at $z=0.25$ (filled markers) in comparison to low-redshift observations 
 (open markers).

 Observationally, supermassive BH spin can be inferred in several ways. The majority of 
 present-day spin constraints come from X-ray reflection spectroscopy -- the analysis 
 of high-energy emission from a hot X-ray corona, reprocessed in a thin accretion disk 
 and `reflected' back to an observer \citep[e.g.][]{Garcia2011,reynolds13,Walton2019,Jiang2019,mallick_high-density_2022,sisk-reynes_evidence_2022}. In this approach, spin is inferred from 
 the observed shape of the Fe K$\alpha$ line at 6.4-6.9~keV, subject to relativistic
 broadening when emitted close to the ISCO \citep[e.g.][]{Fabian1989, Laor1991, Dauser2010} 
 and hence this method is only appropriate
 for sources with prominent iron line signatures. Another important caveat associated
 with reflection spin measurements is their potential vulnerability to confusion between
 K$\alpha$ emission originating from the accretion disk at ISCO, and from high-velocity 
 outflows launched from the disk surface \citep{Parker2022}. Although this degeneracy 
 only affects soft-band X-ray observations and can be broken by including a constraint
 on the Compton hump feature peaking at $\sim 30$~keV 
 \citep[first demonstrated by][]{Risaliti2013}, in practice such measurements require 
 long coordinated exposures on soft- and hard X-ray observatories. For this reason
 only $\sim50$\% of supermassive BH spin measurements to date were obtrained with 
 broadband X-ray coverage \citep[see][for further discussion]{Piotrowska2024}.

 An alternative, albeit less common, approach involves continuum fitting of thermal 
 emission from the AGN disc and relies on the connection between BH spin and expected 
 disc temperature \citep[e.g.][]{Done2013,Capellupo2017}. Since this method does
 not rely on strong iron line emission, it can, in princinple, be applied to a wider 
 range of sources. However, its requirement of generous multiwavelength coverage 
 and, as of yet inaccesible, extreme UV measurement render this approach challenging
 in practice \citep{Capellupo2017}.
 Although different in spirit and precision, both of these methods can only be applied 
 to bright, highly-accreting AGN, as they both rely on the assumption of geometrically 
 thin / optically thick accretion disc models of \citet{Shakura1973} and \citet{Novikov1973} 
 to establish connection between BH spin and accretion disk emission down to the 
 innermost stable circular orbit (ISCO). 

Another approach is to estimate the BH spin from the jet power while making an assumption about the BHs radiative efficiency and Eddington ratio, either directly from observed radio jets or by looking at jetted quasars \citep{soares&nemmen20}. As can be seen by the shape of the markers in Fig. \ref{fig:spin_merger_mass}, BHs in \nh~are predominantly in radio mode (i.e. have $f_{\rm Edd} < 0.01$) at $z=0.25$, with only four BHs in quasar mode. However, the variability of $f_{\rm Edd}$ is high, as discussed in Sec. \ref{sec:gas_accretion}. We explore the impact of this variability, as well as the impact of luminosity on the observable sample of BH spins, further in Section \ref{sec:observability}.
 
 When looking at the distribution of \nh~BH spins, a clear trend emerges with both the current mass of the BH, $M_{\rm BH}$ and with the fraction of this mass gained through BH-BH mergers, $f_{\rm BH,merged}$. We remind the reader that all BHs in \nh~are seeded at $M_{\rm BH,0}= 10^4 \rm \ M_\odot$ with $|a| = 0$. In Fig. \ref{fig:spin_merger_mass}, we identify three dominant regimes for low-redshift BH spins: 
\begin{enumerate}
    \item A gas accretion-driven spin-up for BHs close to their seed mass: After being seeded, BHs are spun up to a maximum spin ($a=0.988$) while doubling to tripling their mass to $2 \times M_{\rm BH,0}$ (here $2 \times 10^4 \ \msun$) through gas accretion. This suggests highly coherent accretion during this early spin-up phase as an initially non-spinning black hole can be maximally spun up by coherently accreting approximately 1.5 times its original mass \citep{Bardeen1970}. As a consequence, the long timescales modelled here mean that our results depend little on the choice of initial spin value for the BHs, unlike shorter studies such as e.g. \citet{cenciBlackHoleSpin2020}. Some BHs continue to grow purely through gas accretion from this point onward (i.e. the fraction of their mass that they gain through mergers, $f_{\rm BH, merged}$ remains zero) until they roughly double their mass again to $\sim 4 \times M_{\rm BH, 0}$.
    \item A merger-induced scattering of spins for low-mass BHs: In the mass range $2-5 \times M_{\rm BH,0}$ (here $ 2 \times 10^4- 5\times 10^4  \ \msun$), the distribution in BH spin becomes much broader due to BH-BH mergers: BHs that have undergone a merger (i.e. that have $f_{\rm BH, merged} >0$) have $0.4 < |a| < 0.988$. Some BHs in this mass range will have acquired all their spin through the BH-BH merger (due to the orbital angular momentum of the merger, the merger-remnant BH of a merger between two non-spinning BHs will have $a=0.69$, \citealp{BertiVolonteri2008}), while others will have at least partially spun up through gas accretion and then had their spin reduced by the merger: highly or maximally spinning BHs statistically decrease their spin during mergers due to the misalignment between the spin of the two merging BHs \citep{BertiVolonteri2008}. \citet{Peirani2024} showed for the same sample of black holes analysed here that during this phase gas accretion is inefficient as the repeated mergers of the host galaxy lead to rapid realignment of the stellar and gas angular momentum with respect to that of the BH.
    \item An accretion-driven evolution for massive BHs: As BHs grow beyond $ 10 \times M_{\rm BH,0}$ (here $10^5 \ \msun$), accretion again dominates the BH mass budget ($f_{\rm BH, merged}$ decreases). BH-BH mergers continue to take place during this late evolution phase (see Fig. \ref{fig:spin_time_mostmassive} below) but the mass ratio of mergers decreases as the likelihood of encountering an equally massive BHs becomes smaller. As a result, BHs assemble their mass predominantly through gas accretion. Early on, the scatter remains high, while for the most massive BHs in \nh, spin magnitudes again increase towards maximally spinning.
\end{enumerate}

While our sample of massive BHs is too small to draw firm conclusions on the spin distribution of massive BHs in the present day Universe, those that lie within the same BH mass range as the BHs observed in the local Universe also show absolute spin magnitudes, $|a|$, in good agreement with the observations as can be seen in Fig. \ref{fig:spin_merger_mass}.

\begin{figure*}
    \centering
    \includegraphics[width=\textwidth]{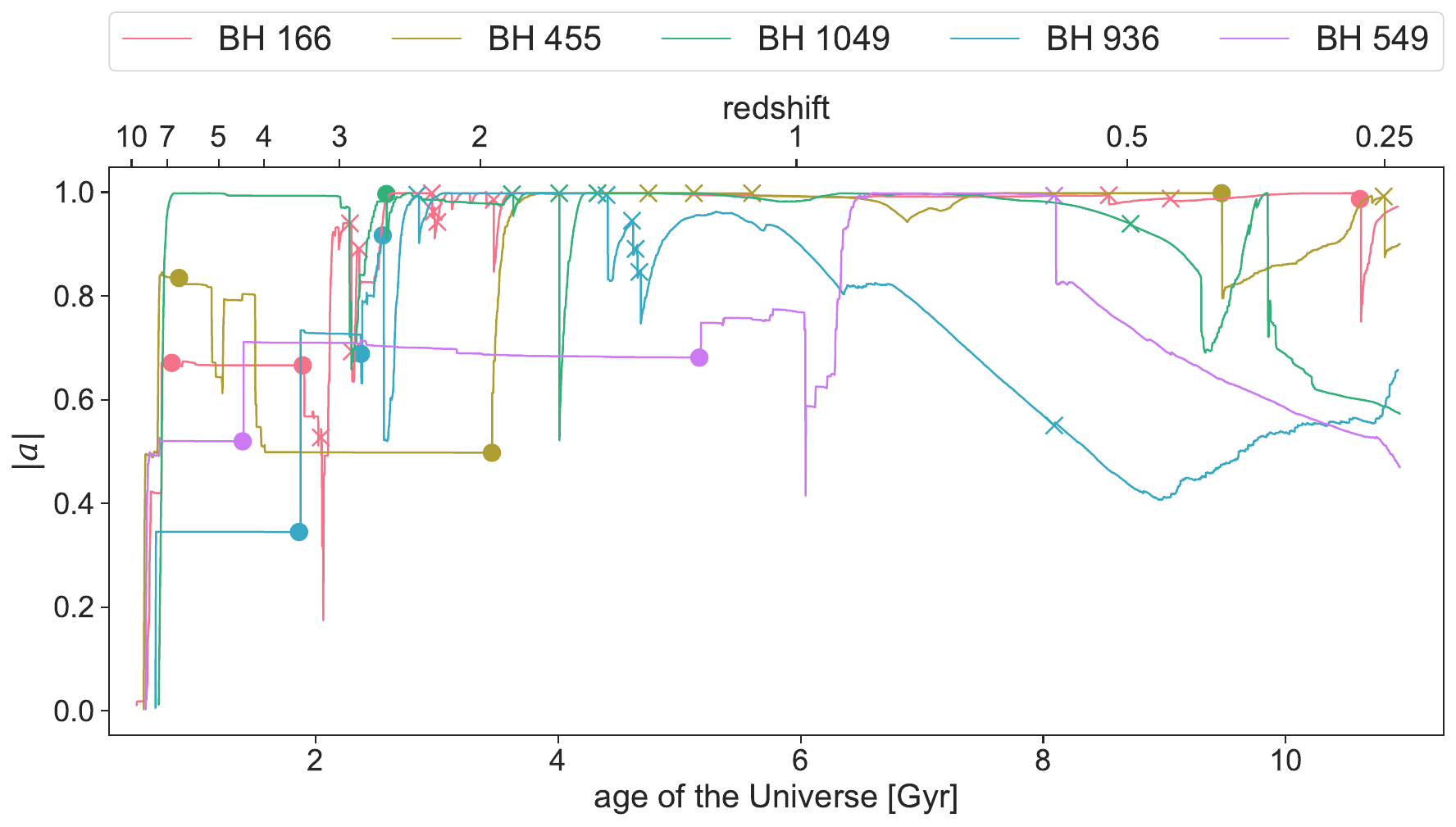}
    \caption{Timeseries for the absolute spin magnitude $|a|$ for the five most-massive BHs in \nh~ (equivalent to a mass cut of $M_{\rm BH} > 2 \times 10^5 \ \msun$). Circular markers denote the timing of major BH-BH mergers (mass ratio $> 1:4$), while crosses denote minor mergers.}
    \label{fig:spin_time_mostmassive}
\end{figure*}

Fig. \ref{fig:spin_time_mostmassive} shows how the spin of the five most massive BHs in \nh~evolves over time. The same three trends as discussed in Fig. \ref{fig:spin_merger_mass} qualitatively emerge from these timeseries: first, there is an early spin-up of newly formed BHs to intermediate spin magnitudes ($0.3 < |a| < 0.9$). For the next few Gyr of BH evolution, BH-BH mergers can produce large jumps in $|a|$, often but not consistently reducing it. At late times ($z<2$) most mergers have a modest impact on $|a|$ and gradual change through gas accretion dominates the evolution of the BH spin magnitude. This is discussed further in Sec. \ref{sec:mergers}.

\begin{figure*}
    \centering
    \includegraphics[width=0.95\textwidth]{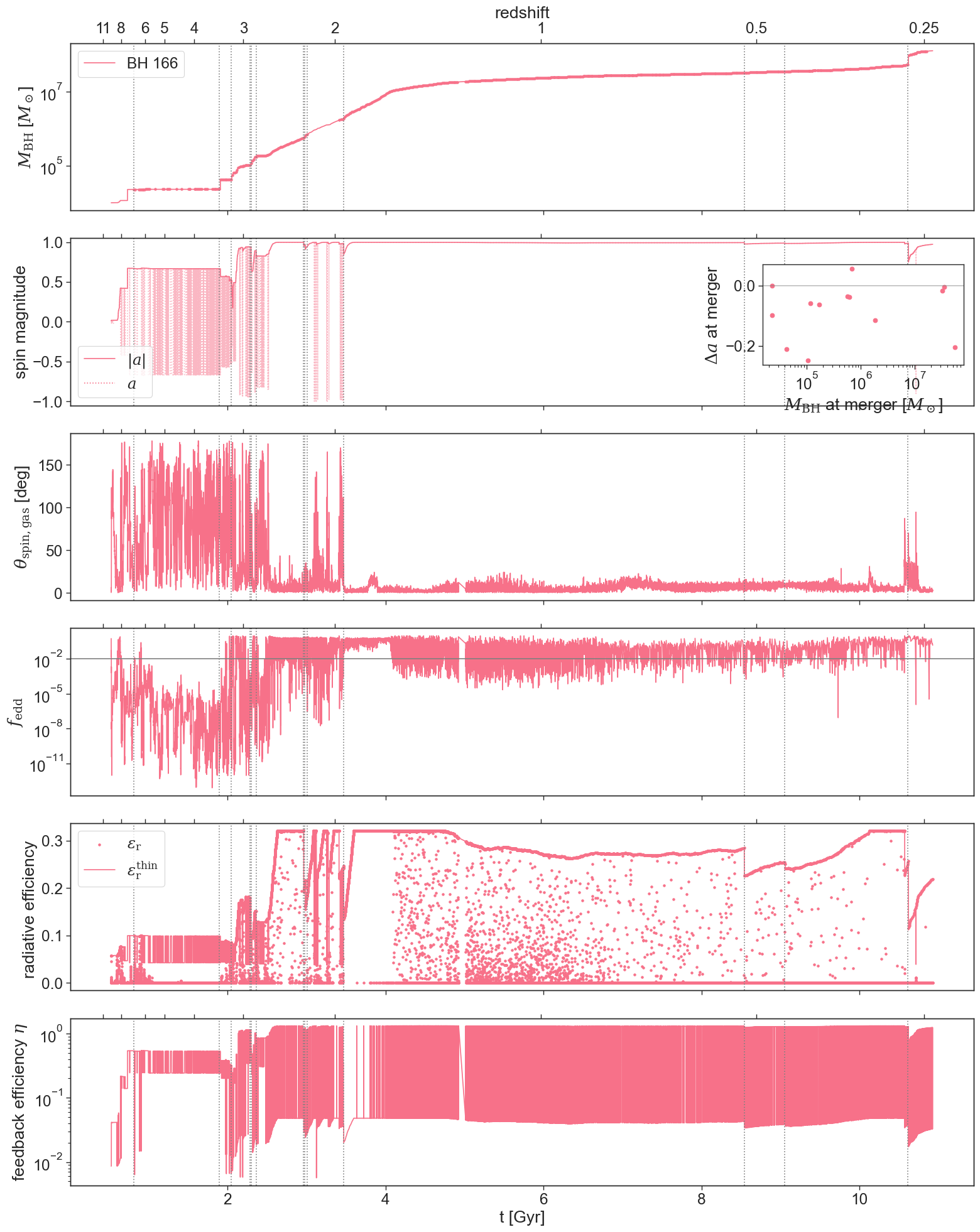}
    \caption{Timeseries for BH 166 of (from top to bottom) BH mass $M_{\rm BH}$, absolute and intrinsic BH spin magnitude $|a|$ and $a$, angle between accreted gas and BH spin $\theta_{\rm spin,gas}$, Eddington ratio $f_{\rm Edd}$, spin-dependent radiative efficiency $\varepsilon_{\rm r}$ and BH feedback coupling efficiency $\eta$. Grey vertical lines mark the time of BH-BH mergers with a minimum mass ratio of $q=0.1$.}
    \label{fig:bh166}
\end{figure*}

\begin{figure*}
    \centering
    \includegraphics[width=0.95\textwidth]{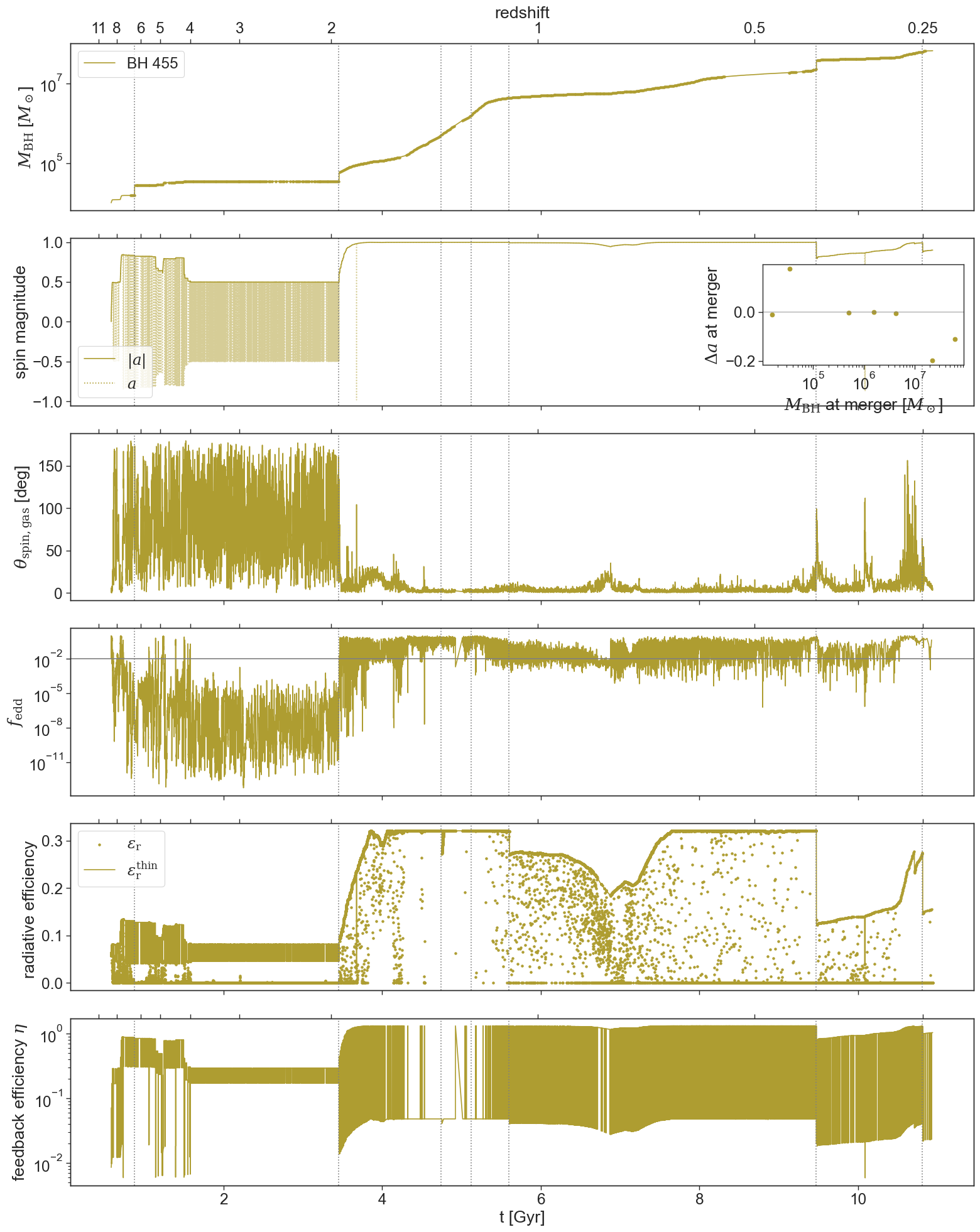}
    \caption{Timeseries for BH 455. See Fig. \ref{fig:bh166} for details.}
    \label{fig:bh455}
\end{figure*}

\begin{figure*}
    \centering
    \includegraphics[width=0.95\textwidth]{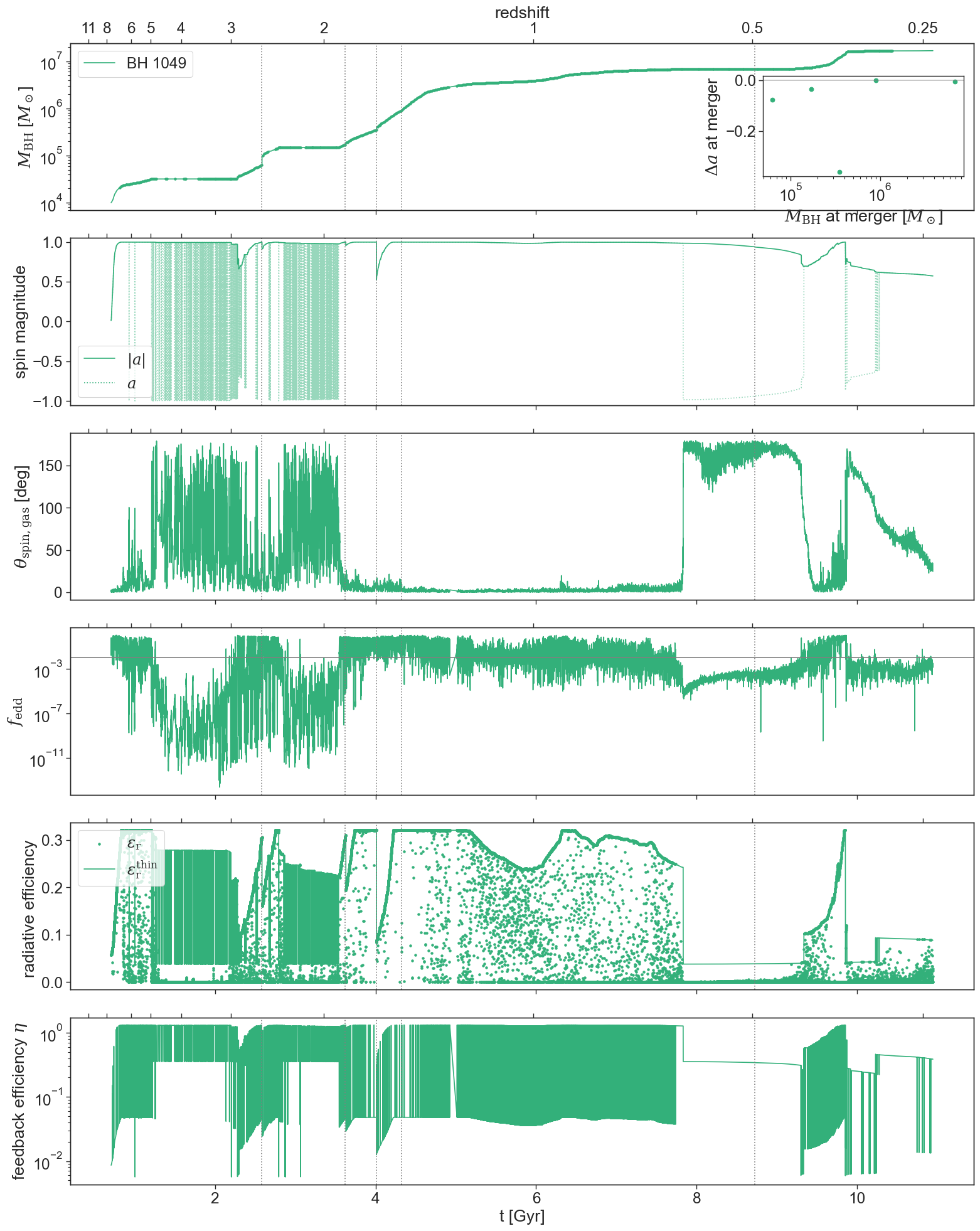}
    \caption{Timeseries for BH 1049. See Fig. \ref{fig:bh166} for details.}
    \label{fig:bh1049}
\end{figure*}

\begin{figure*}
    \centering
    \includegraphics[width=0.95\textwidth]{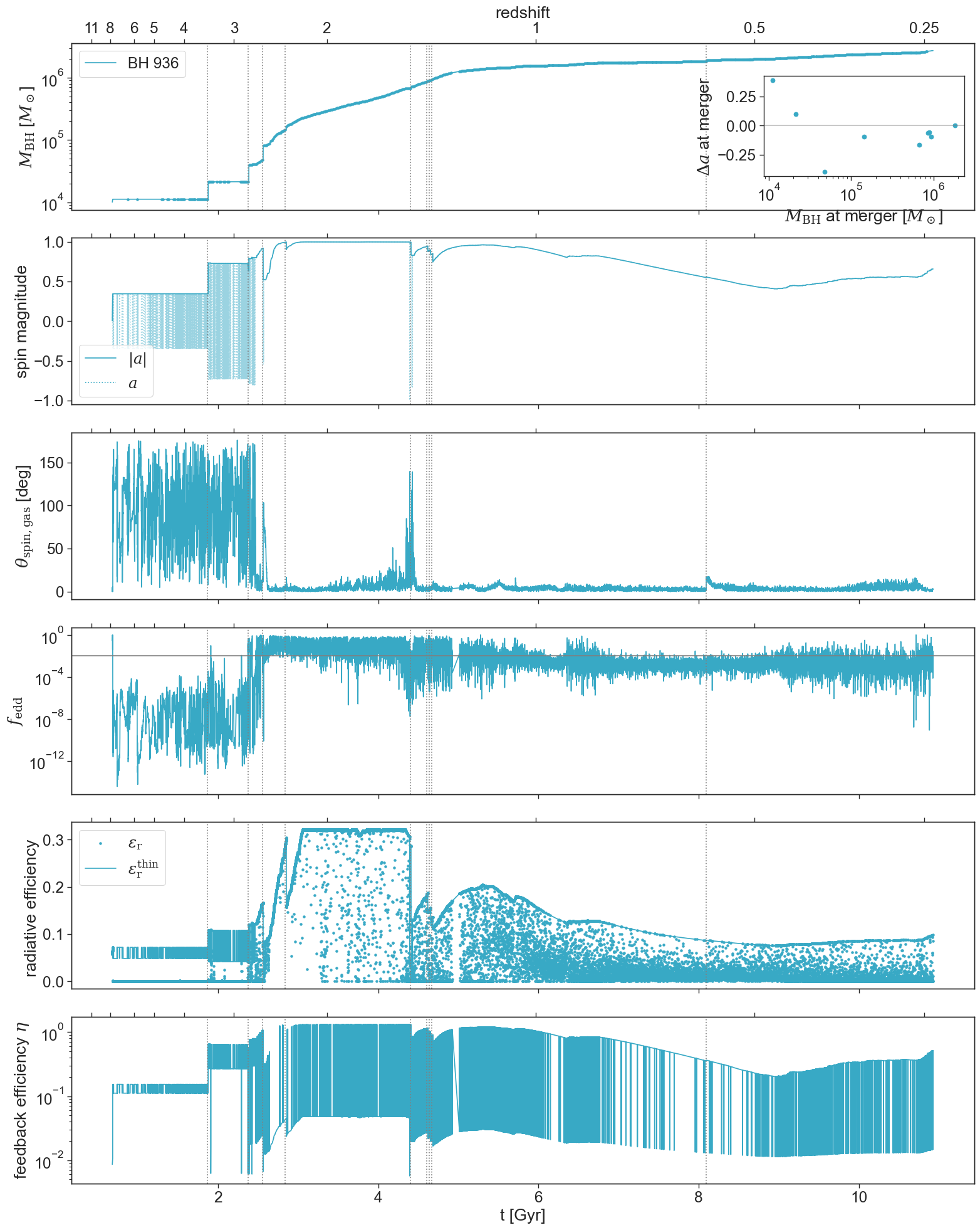}
    \caption{Timeseries for BH 936. See Fig. \ref{fig:bh166} for details.}
    \label{fig:bh936}
\end{figure*}

\begin{figure*}
    \centering
    \includegraphics[width=0.95\textwidth]{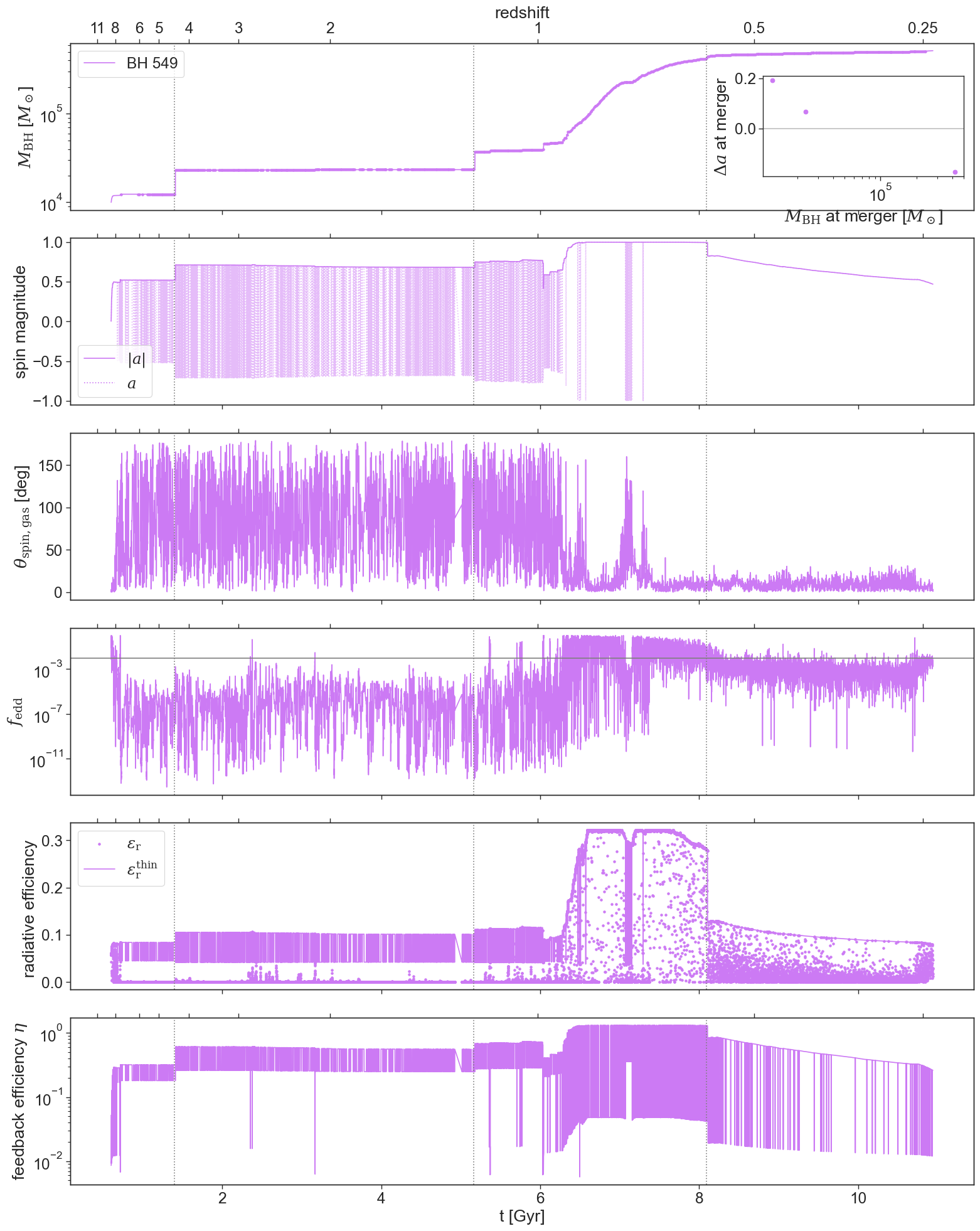}
    \caption{Timeseries for BH 549. See Fig. \ref{fig:bh166} for details.}
    \label{fig:bh549}
\end{figure*}

The time evolution of these five BHs is explored further in Fig. \ref{fig:bh166} to \ref{fig:bh549}, which show (from top to bottom): BH mass $M_{\rm BH}$, absolute and intrinsic spin magnitude $a$, the angle between accreted gas and BH spin $\theta_{\rm spin,gas}$, the Eddington ratio $f_{\rm Edd}$, the spin-dependent radiative efficiency $\varepsilon_{\rm r}$ and the feedback coupling efficiency $\eta$ for the five most massive BHs in \nh~ at $z=0.25$. The angular momentum of the accreted gas is measured as described in Section \ref{sec:BHacc}. A few quantitative trends emerge: BHs tend to have a chaotic, inefficient accretion phase early on in their evolution, which lasts anywhere from 2 to 6 Gyr. After this, accretion becomes more coherent and efficient, growing BHs in mass and spinning them up. As discussed in more detail in Section \ref{sec:gas_accretion}, this time evolution is at least partially driven by the mass evolution of the BH's host galaxy.

At late times, two separate trends can be seen, linked to the accretion states of BHs: BHs predominantly in ``quasar'' mode (BH 166 in Fig. \ref{fig:bh166}, 455 in Fig. \ref{fig:bh455}), which have an average $f_{\rm Edd}=0.03-0.6$ over the last 3 Gyr of simulation are spun up by gas accretion, even after temporary spin-down following mergers (see Fig. \ref{fig:spin_time_mostmassive}). BHs growing less efficiently (here BH 936 in Fig. \ref{fig:bh936} \& 549 in Fig. \ref{fig:bh549}, which have an average $f_{\rm Edd}\sim 0.004$ over the last 3 Gyr. We note that BH 936 spins up again at late times.) are spun-down on Gyr timescales as spin energy is extracted to power the jets. BH 1049 shows an interesting mix of both behaviours, with a long spin-down phase followed by a rapid spin-up phase, followed by more spin-down. We discuss the impact of accretion mode on spin evolution further in Section \ref{sec:gas_accretion}.

Overall, our results agree with previous simulation studies on the long-term spin evolution of massive BHs using cosmological simulations by \citet{Dubois2014} and \citet{Bustamante2019} that showed that the spin of massive BHs is generally high and can remain roughly constant on Gyr timescales. Unlike these works, we find examples of non-maximally-spinning BHs in the mass-range $10^6 - 10^ 8 \rm \ M_\odot$, which are spun down by coherent but inefficient accretion, in which BH spin energy is extracted to drive BH jets. This effect was not modelled in the previous two works. Unfortunately, our sample is too small to draw firm conclusions on how this affects the total spin distribution, but the two examples reported here provide intriguing hints, together with recent observations by \citet{mallick_high-density_2022}, that BH spin-down by jets could play an important role in broadening the distribution of BH spins in the mass range $10^6 - 10^8 \rm M_\odot.$ \citet{Sala2023} reported that BH maximally spin up as they double their seed mass, but then report an intermediate period of high spin in the mass range $10^6- 10^7 M_\odot$ before a mix of BH mergers and spin-down due to chaotic accretion increases scatter in spin at late times for BHs in the mass range $10^7 - 10^{10} M_\odot$ even without jet spindown, as also suggested by \citet{Fiacconi2018}.

\subsection{The impact of black hole mergers on black hole spin}
\label{sec:mergers}

\begin{figure*}
    \centering
    \includegraphics[width=\textwidth]{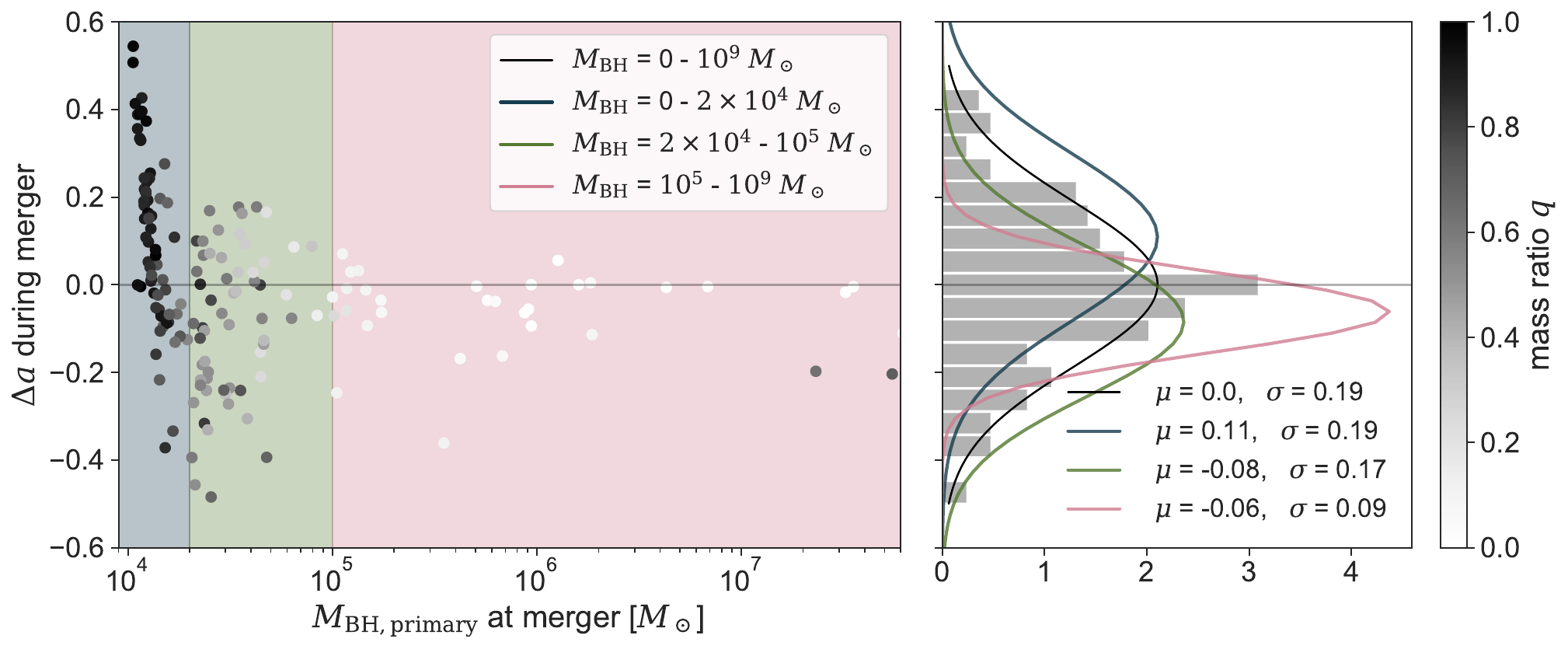}
    \caption{Left: Distribution of BH spin changes $\Delta a$ during BH-BH mergers versus the primary BH mass at the time of merger, $M_{\rm BH,primary}$ (left). Right: Gaussian fit to the distribution for the total sample (black) and the three subsamples (coloured).}
    \label{fig:merger_spinchange}
\end{figure*}

The decreasing importance of mergers with BH mass in the mass range explored here can be seen more quantitatively in Fig. \ref{fig:merger_spinchange}, which explicitly shows the distribution of the changes in spin magnitude $\Delta a$ as a function of the mass of the primary BH for all BH-BH mergers in \nh. This plot shows a sample of stacked BH mergers that occur at different redshifts throughout the simulation. The same three phases as in Fig. \ref{fig:spin_merger_mass} can be broadly identified:
\begin{enumerate}
    \item First BH mergers, during which the mass of the primary, $M_{\rm BH,primary} < 2 \times M_{\rm BH,0}$ (blue background). Such BHs are still in the process of spinning up through early gas accretion. BHs very close to the seed mass have very low $|a|$ pre-merger (see Fig. \ref{fig:spin_merger_mass}) and are therefore statistically spun up by mergers ($\Delta a > 0$). The opposite is true for BHs that have grown through gas accretion up to twice their original mass: they tend to be highly spinning pre-merger (see Fig. \ref{fig:spin_merger_mass}) and are therefore likely to be spun down by the first merger ($\Delta a < 0$). However, mergers of non-grown BHs are more common, so the average merger during this early phase will spin BHs up (see right-hand panel of Fig. \ref{fig:merger_spinchange}). By definition, BH mergers for BHs close to the seed mass tend to have mass ratios close to unity.
    \item An intermediate phase, where the primary is in the mass range $2 - 10 \times M_{\rm BH,0}$ (green background). These are BHs that have already undergone either significant gas accretion or previous BH-BH mergers and will therefore have a range of spin values and orientations pre-merger. Spin changes during this period can be both positive and negative, but on average reduce the BH spin (see right-hand panel of Fig. \ref{fig:merger_spinchange}). Mergers during this phase tend to remain major mergers (mass ratio $q>1:4$) but decrease with increasing primary mass.
    \item The long-term evolution of massive BHs (red background):  Mergers continue to take place and continue to on average decrease the spin magnitude, but the average spin change per merger becomes significantly smaller as the average mass ratio $q$ between merging BHs decreases. The two major mergers that occur for massive BHs in \nh~also show that larger spin-changes are possible for massive BHs as long as the merger ratio is sufficiently high.
\end{enumerate}

We note that having such clean mass transitions between the regimes is partially due to the uniform seed mass of $M_{\rm BH,0} = 10^4 \rm \msun$ and initial spin ($a=0$) for all BHs in \nh. One consequence of a uniform initial mass function of BH seed masses is that all early mergers are by necessity major mergers. For a more diverse initial mass function of $M_{\rm BH,0}$ we would expect early mergers to continue to play an important role in increasing the scatter in BH spins for BHs of $10^5 M_\odot$, but the mass transitions would become less clear as early mergers would encompass a range of minor and major mergers with different mass ratios. 

\subsection{The impact of gas accretion on black hole spin}
\label{sec:gas_accretion}

Understanding how gas accretion influences the long-term spin evolution of BHs is more complex because the efficiency with which angular momentum is transferred from the gas to the BH (i.e. with which the BH is spun up or down by gas accretion) depends on 1) the existing spin of the BH 2) the long-term alignment between the accreted gas angular momentum vector and the spin vector of the BH, here called $\theta_{\rm gas, BH}$ and 3) the Eddington ratio $f_{\rm Edd}$ as it determines the BH feedback mode. 

BHs with $f_{\rm Edd}>0.01$ are assumed to be in `quasar' mode, during which gas accretion will always spin BHs up if the angular momentum of the accreted gas is aligned with the BH spin, with the spin-up rate depending on the current spin of the BH. BHs with $f_{\rm Edd}<0.01$ are assumed to be in `jet' mode where the combined action of angular momentum transferred to the BH and BH spin energy extracted to power the jet means that $|a| \rightarrow 0$ over time \citep[see Sec. 2.5.2 in][for details]{Dubois2021}.

Here we assume that the angular momentum direction of the accretion disc during any accretion event is equal to the accreted gas' angular momentum direction. Long periods of coherent gas flows on galactic scales allow BH spin vectors to realign themselves with the angular momentum vector of the inflowing gas, which then creates ideal conditions for efficiently spinning BHs up. By contrast, if the angular momentum of the inflowing gas frequently changes, due to chaotic clumpy accretion onto the BH, we expect the spin magnitude $a$ to decrease if accretion is sufficiently efficient as retrograde accretion carries more specific angular momentum than pro-grade accretion.

The interplay between $\theta_{\rm gas,BH}$, $f_{\rm Edd}$ and $a$ is complex, as can be seen by the evolutionary histories of the most massive BHs shown in Sec. \ref{sec:spin}.  Qualitatively, all BHs undergo an early period where $f_{\rm Edd}$ is generally low and $\theta_{\rm gas,BH}$ varies rapidly. Early in their evolution, when galaxies are regulated by stellar feedback \citep{Dubois2015,Habouzit2017,Bower2017,Angles-Alcazar2017,trebitschetal17}, they have highly disturbed morphologies and lack coherent gas inflows into the centre. As a result, gas falling onto the BHs has rapidly changing angular momentum, and $a$  oscillates between being positive and negative, as the sign of $a$ is measured with respect to the angular momentum of gas that is currently being accreted. This stochasticity could partially be due to the resolution of the simulation. It has been shown that resolving the gas flows around BHs in more detail can turn accretion from chaotic to episodic as a nuclear disc around the BH becomes resolved \citep{Beckmann2019,Hopkins2023}, or make accretion more stochastic as the inner disc fragments \citep{Levine2010} depending on the conditions of the disc.

This phase can last anywhere from 1 to 3 Gyr, depending on the BH, and is also a time of frequent BH-BH mergers with sufficiently high mass ratios to significantly boost the BH mass and influence the absolute spin magnitude $|a|$. For most BHs, eventually $\theta_{\rm gas,BH}$ settles down and $f_{\rm Edd}$ increases as the BH enters an accretion-driven regime where it grows steadily in mass. Then follows a period of coherent accretion onto the BH where BH spin is well-aligned with that of the gas in the centre of the galaxy, which we assume flows coherently all the way to the accretion disc. Mergers continue to occur but as merger ratios drop, the impact on the BH mass and spin evolution diminishes. We note that $\theta_{\rm spin,gas}$ only measures the relative angle between the BH spin and the accreted gas angular momentum. It does not quantify the direction of either gas or BH with respect to the host galaxy or large-scale cosmological environment. $\theta_{\rm spin,gas} \sim 0$ means that the angular momentum of accreted gas is changing slowly enough for the BH to continue aligning with it, not that the BH spin vector (and accreted gas angular momentum vector) is fixed in 3D space during this period of time.

\begin{figure*}
    \centering
    \includegraphics[width=\textwidth]{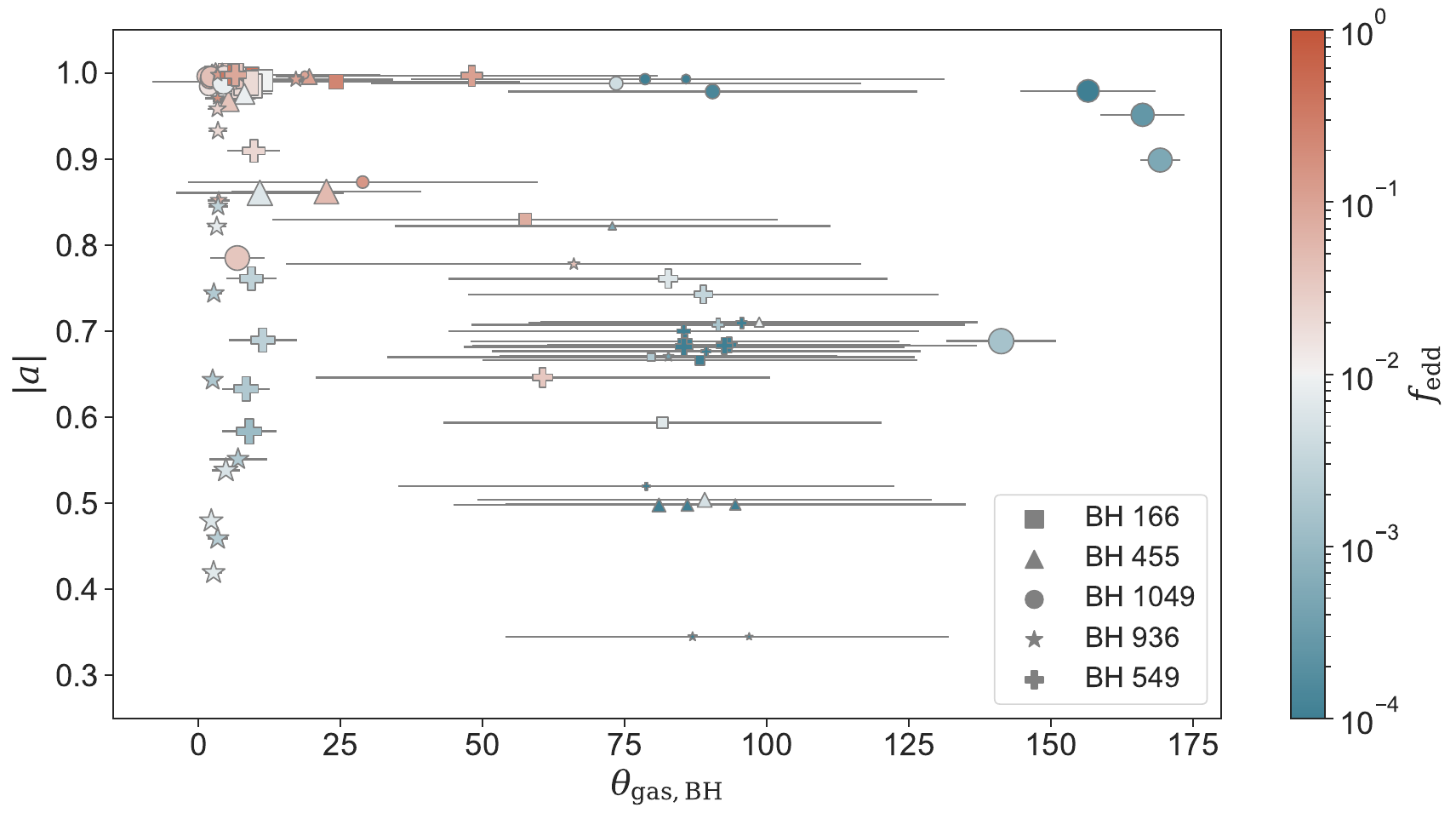}
    \caption{Distribution of the absolute spin magnitude $|a|$ versus the angle between accreted gas angular momentum and BH spin vector, $\theta_{\rm gas, BH}$, for the five most massive BHs in \nh~ (the same ones as shown in Fig. \ref{fig:spin_time_mostmassive}). BHs are sampled every 0.5 Gyr, with markers denoting the average $|a|$ and $\theta_{\rm gas, BH}$ over $\Delta t = 100 \rm \ Myr$. Each marker is colour-coded by the average $f_{\rm Edd}$ over the same period. Grey errorbars show the variance of $\theta_{\rm gas, BH}$ over each 100 \rm \ Myr time interval. Small errorbars indicate a period of high coherence in the accreted gas, while large errorbars are a sign of chaotic accretion. Marker size encodes cosmic time, with smaller markers presenting earlier points in a BHs evolution history.}
    \label{fig:coherence}
\end{figure*}

In Fig. \ref{fig:coherence} we explore the impact of gas accretion on BH spin evolution more quantitatively. In this plot, we compute the average absolute spin magnitude $|a|$ and the average angle between gas angular momentum and BH spin vector every $0.5\,\rm Gyr$ over intervals of $100 \,\rm Myr$ for the five most massive BHs in \nh. Markers show the average value of $|a|$ and $\theta_{\rm gas, BH}$ over each interval of  $100 \,\rm Myr$, while grey errorbars denote the variance in $\theta_{\rm gas, BH}$ over the same time interval. Colour-coding indicates the average $f_{\rm Edd}$ for each interval, and marker size denotes cosmic time, with smaller markers presenting earlier sampling times. As can be seen in Fig. \ref{fig:coherence}, early on all BHs shown here undergo chaotic accretion (error bars in $\theta_{\rm gas, BH}$ are large, which shows a lack of coherence in the accreted gas). During this time, BH accretion is very inefficient ($f_{\rm Edd}$ is low). \citet{Peirani2024} showed that this chaotic phase is driven by a rapid reorientation of the stellar angular momentum, rather than by a reorientation of the BH spin. Eventually either the BH undergoes a period of efficient, more coherent accretion (small, reddish markers) or the BH is spun up by an early merger (see Sec. \ref{sec:mergers}).

At later times, as the host galaxy grows in mass and settles, the coherence of the accreted gas settles down (i.e. errorbars in $\theta_{\rm gas, BH}$ become small) and all BHs studied here undergo a period of efficient accretion. \citet{Peirani2024} showed that this late transition occurs around a host galaxy mass of $5 \times 10^9 - 5 \times 10^{10} M_\odot$.

During this time, gas accretion keeps BH spins high, as the high $f_{\rm Edd}$ put BHs in quasar feedback mode,  allowing them to be spun even when BH mergers briefly decrease their spin magnitudes. Examples of such BHs are BH 166 and BH 455, which both show persistent high spin magnitudes at late times (see also Fig. \ref{fig:spin_time_mostmassive}). For other BHs, $f_{\rm Edd}$ eventually drops, pushing BHs into jet feedback mode. As a result, such BHs are slowly spun down over Gyr timescales, as can be seen for BHs 936 and 549 in Fig. \ref{fig:spin_time_mostmassive}). Interestingly, BH 1049 undergoes a period of persistent anti-aligned accretion, also evident in its timeseries in Fig. \ref{fig:bh1049} around $z=0.5$ that causes slow spindown.

\begin{figure}
    \centering
    \includegraphics[width=\columnwidth]{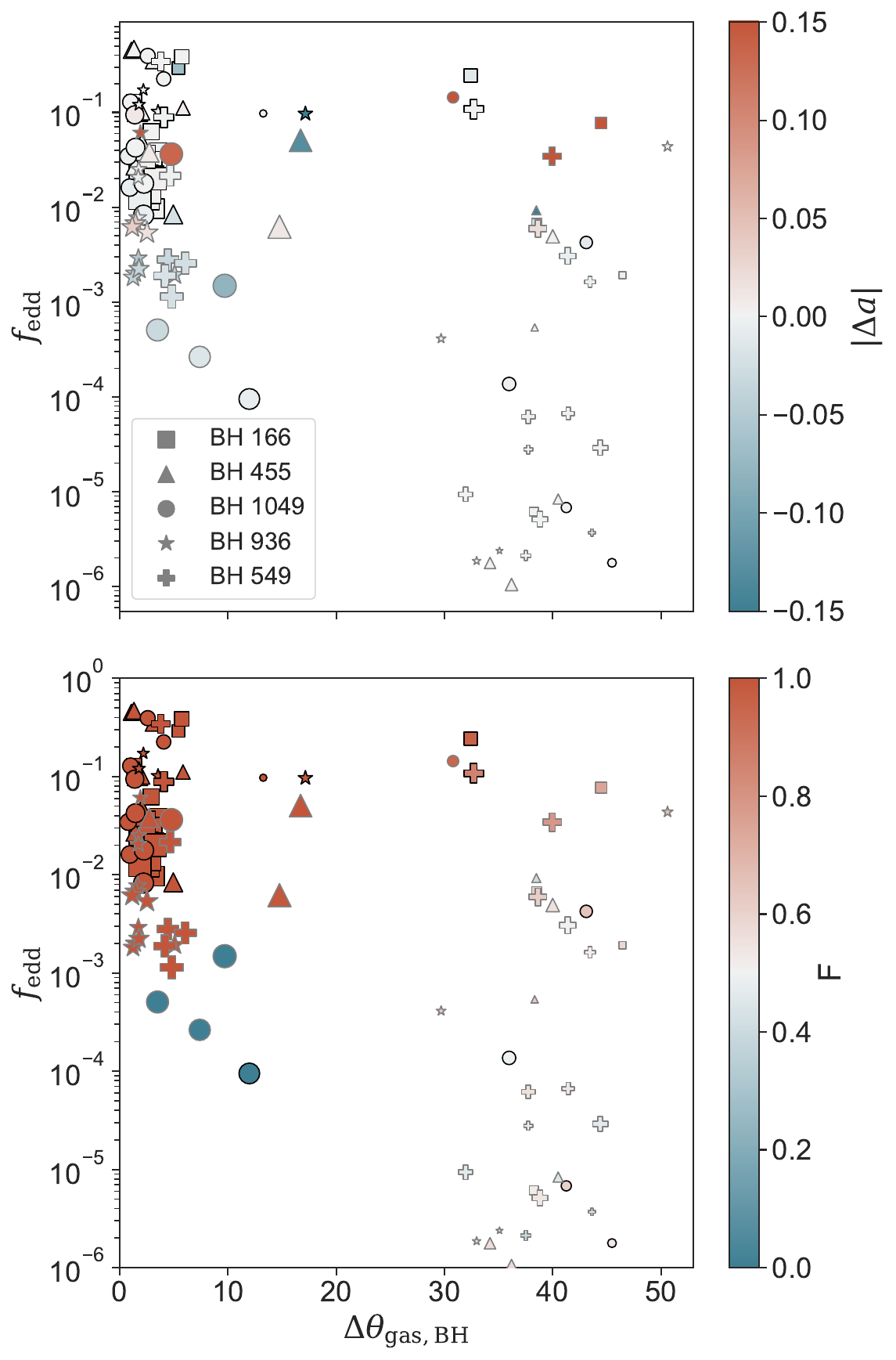}
    \caption{Distribution of the variance of $\theta_{\rm gas, BH}$ versus the average Eddington ratio $f_{\rm edd}$ for the 5 most massive BHs in \nh~ sampled every 0.5 Gyr over 100 Myr intervals, as in Fig. \ref{fig:coherence}. Black outlines show BHs with average spin magnitude $|a|> 0.95 $ i.e. almost maximally spinning. Data points are colour-coded by the total spin change due to accretion $\Delta a_{\rm acc}$ during each interval (top panel) and by the anisotropy parameter $F$ (bottom panel, see \citealp{Dotti2013}). The impact of mergers has been removed from $\Delta a_{\rm acc}$.}
    \label{fig:chaotic}
\end{figure}

We investigate the impact of (in)coherence further in Fig. \ref{fig:chaotic} which is based on the same data points as Fig. \ref{fig:coherence}. We now show the variance $\Delta \theta_{\rm BH,gas}$ on the x-axis (shown as the length of the errorbars in Fig. \ref{fig:coherence}) versus the Eddington ratio $f_{\rm edd}$. As can be seen in the top panel of Fig. \ref{fig:chaotic}, many BHs undergoing highly incoherent accretion (high $\Delta \theta_{\rm BH,gas}$) show little spin change $\Delta a$ because they are growing inefficiently (low $f_{\rm edd}$). Contrary to expectation, we find that BHs that are undergoing efficient incoherent accretion (high $f_{\rm edd}$) are actually spun up ($\Delta a >0$). We quantify the anisotropy in the accreted gas using the anisotropy parameter $F$, following the analysis in \cite{Dotti2013}. $F$ measures the fraction of accretion events for which the dot product between BH spin and gas angular momentum is positive. Truly random accretion has $F=0.5$ while ($F<0.5$) $F>0.5$  suggests on average (anti-)aligned accretion. We find that for our BHs, even during their highly incoherent accretion phases, $F$ remains sufficiently high to cause long-term spinup. The few BHs that show high $f_{\rm edd}$ and low $\Delta a$ are almost maximally spinning already (black outline), so cannot be spun up further. Note the anti-aligned period of accretion of BH 1049 again which leads to a net spin-down during this time.

Overall we find little evidence for an early suppression of spin-up due to chaotic accretion in the early Universe or the late spin-down of massive BHs due to chaotic accretion \citep{Volonteri2013,Bustamante2019,griffin_evolution_2019}. This is unlike previous results reported by \citet{IzquierdoVillalba2020}, \citet{Bustamante2019} and \citet{Sesana2014} who stress the importance of spin-down through chaotic accretion to produce the diverse distribution of BH spins observed in the local Universe.
Instead, in \nh~ a qualitatively similar spin-down is produced by extracting BH spin to drive jets for some massive BHs, an effect that was not included in  the works cited above. This could mean that previous works over-estimated the impact of chaotic accretion to compensate for the lack of jet-induced spin-down. Alternatively it might show that our assumption of the gas at BH accretion scales inheriting the angular momentum from gas at resolution scales makes efficient accretion artificially coherent, reducing spin-down via chaotic accretion. A combination of both effects is also possible.

In conclusion, through a combination of efficient thin disc accretion, non-negligible periods of anti-aligned accretion and jet-induced spin-down, \nh~predicts a diverse population of BHs at low redshift with the potential for a large dispersion in spin values. Unfortunately, our sample is too small, and does not reach very high BH masses, to robustly quantify the distribution of massive BH spins in the local Universe or study the impact of environment driving this diversity.

We find little evidence for spin-down through chaotic accretion even at early times, as the chaotic accretion episodes we observe early-on in the evolution of most of our massive BHs is either too inefficient to significantly influence BH spin or leads to a new spin-up due to persistent coherence during such episodes.

\subsection{The impact of black hole spin on feedback efficiencies}
\label{sec:radiative_efficiency}

\begin{figure*}
    \centering
    \includegraphics[width=\textwidth]{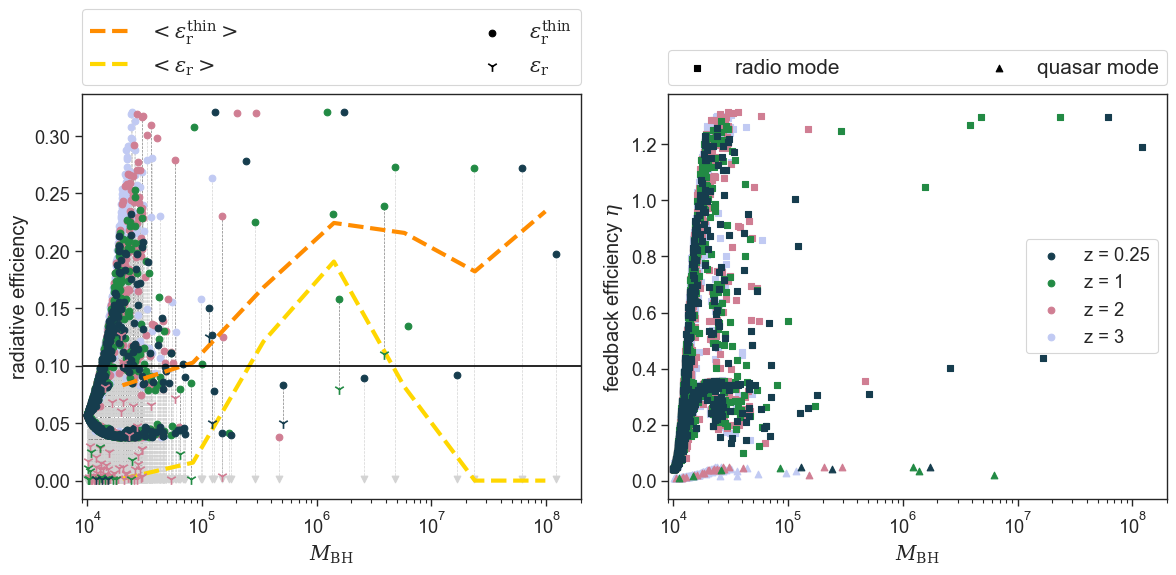}
    \caption{Distribution of the spin-dependent radiative efficiency of a thin disc $\varepsilon_{\rm r}^{\rm thin}$ and the effective radiative efficiency $\varepsilon_{\rm r}$ (left) and the feedback efficiency $\eta$ (right) as a function of BH mass for BHs in \nh~ at redshifts $z=3,2,1$ and $z=0.25$. For BHs in quasar mode, where $\varepsilon_{\rm r} = \varepsilon_{\rm r}^{\rm thin}$, only one data point is shown. For BHs in radio mode, where $\varepsilon_{\rm r}= f_{\rm att} \varepsilon_{\rm r}^{\rm thin}$, any BHs with $\varepsilon_{\rm r} < 10^{-3}$ are shown as limits. The dashed lines in the left plot show the average $\varepsilon_{\rm r}^{\rm thin}$ and $\varepsilon_{\rm r}$ as a function of mass across all redshifts.}
    \label{fig:epsilon}
\end{figure*}

One important consequence of following BH spin is that it changes both the radiative efficiency $\varepsilon_r$, which controls mass growth and luminosity of the BH, and the feedback efficiency $\eta$, which controls how much of the feedback energy of the BH couples to the ISM close to the BH (see Sec. \ref{sec:BHfeed} for details) Both potentially have important consequences for BH and galaxy coevolution, as well as their observability.

In the absence of spin information, most existing galaxy evolution simulations use a fixed efficiency of $\varepsilon_{\rm r} =0.1$ as radiative efficiency, often attenuated by a factor calibrated on low-redshift observation for the feedback efficiency. In our model, the radiative efficiency $\varepsilon_{\rm r}$, and feedback efficiency $\eta$ are both based on analytic models that take spin evolution into account (except for a factor of 0.15 in the quasar feedback efficiency used for calibration, see Sec. \ref{sec:BHfeed} for details).

In Fig. \ref{fig:epsilon} we show the spin-based distribution of $\varepsilon_{\rm r}^{\rm thin}$ and $\varepsilon_{\rm r}$ (where $\varepsilon_{\rm r} = f_{\rm att} \varepsilon_{\rm r}^{\rm thin}$ includes the attenuation for BHs in radio mode, see Sec. \ref{sec:BHfeed}) for all BHs at four different redshifts (left), and the feedback efficiency $\eta$ (right). Both are highly variable (see the bottom two panels of Fig. \ref{fig:bh166} to \ref{fig:bh549} for examples). Fig. \ref{fig:epsilon} shows that BHs close to the seed mass have a wide range of $\varepsilon_{\rm r}^{\rm thin}$. The bimodal nature of the distribution is driven by BHs that are aligned ($a>0$, upper branch) and anti-aligned ($a<0$, lower branch) with the angular momentum of gas that they are accreting. While some BHs in the mass range $10^5 - 10^6 \,\rm M_\odot$ also have thin discs radiative efficiency $\epsilon_{\rm r}^{\rm thin}$ below the classic value (horizontal line), most massive BHs have $\varepsilon_{\rm r}^{\rm thin} > 0.1$. The average value for all BHs with masses $M_{\rm BH} > 10^5 \rm \ M_\odot$ is $<\varepsilon_{\rm r}^{\rm thin}> = 0.19$. This means that massive BHs in quasar mode in \nh~will shine 2-3 times as brightly at a given accretion rate as previously assumed, which could also have important consequences when using the luminosity to derive the spin value of BHs \citep[see e.g.][who assume an efficiency of $0.15$]{soares&nemmen20}. However, attenuating the radiative efficiency by a factor of $f_{\rm att}$ for BHs in radio mode, to account for the fact that such BHs have radiatively inefficient accretion discs, has a significantly bigger impact on the average radiative efficiency than using a spin-driven $\varepsilon_{\rm r}^{\rm thin}$, as can be seen by the fact that most AGN in jet mode, even those with high masses, end up with an effective radiative efficiency $\epsilon_{\rm r} < 10^{-3}$ (shown as grey upper limits). As a result, the average $\epsilon_{\rm r}$ for all AGN above $10^{5} \rm \ M_\odot$ is only $<\varepsilon_{\rm r}> = 0.1$, but the distribution is very double-peaked between those in quasar mode (which typically have $\varepsilon_{\rm r} > 0.25$) and those in radio mode (which typically have $\varepsilon_{\rm r} < 0.1$). From this, we conclude that the high spin values of massive BHs will make those with thin accretion discs easier to observe but delay their mass growth. As most BH accumulate most of their accreted mass in quasar mode, this could be the reason why central BHs in massive galaxies in \nh~ tend to lie towards the lower end of the observed distribution of BH masses \citep[see][]{Beckmann2023}. For BHs with thick discs, observing their disc emission remains challenging, even with high spins, but mass growth is more efficient for such BH as almost all the mass flowing onto the accretion disc from large scales can be accreted due to their low $\epsilon_{\rm r}$.

Following the spin evolution of the BH also allows us to more accurately estimate the amount of feedback energy from the BH that couples to the ISM. This quantity is encoded in the feedback efficiency $\eta$, shown in the right-hand panel of Fig. \ref{fig:epsilon} as a function of BH mass. Here the impact of spin on the distribution is even more striking. When taking the spin into account, BHs in the mass range $> 10^{5} \ M_\odot$ now have an average $<\eta> = 0.04$, roughly 3 times higher than the fiducial value, and $<\eta> = 0.8$ in jet mode which is 8 times higher than the previous assumptions based on a fixed spin and hence fixed radiative efficiency of $\epsilon_{\rm r}^{\rm thin} = 0.1$. So tracking spin evolution means that all galaxies hosting massive BHs experience significantly more feedback energy for a fixed accretion rate than previously assumed.

The overall expected impact on galaxy evolution from an increased feedback efficiency is difficult to estimate. First, the effect is not uniform across a galaxy's evolution: only as BHs spin up over time does their feedback efficiency increase. Secondly, the total amount of feedback energy experienced by a galaxy remains determined predominantly by variations in the accretion rate (see Fig. \ref{fig:bh166} to \ref{fig:bh549}), rather than the feedback efficiency. As BH feedback self-regulates, the overall amount of energy experienced by a given galaxy might not be so different for a fixed-spin and a spin-driven model.

\begin{figure*}
    \centering
    \includegraphics[width=\textwidth]{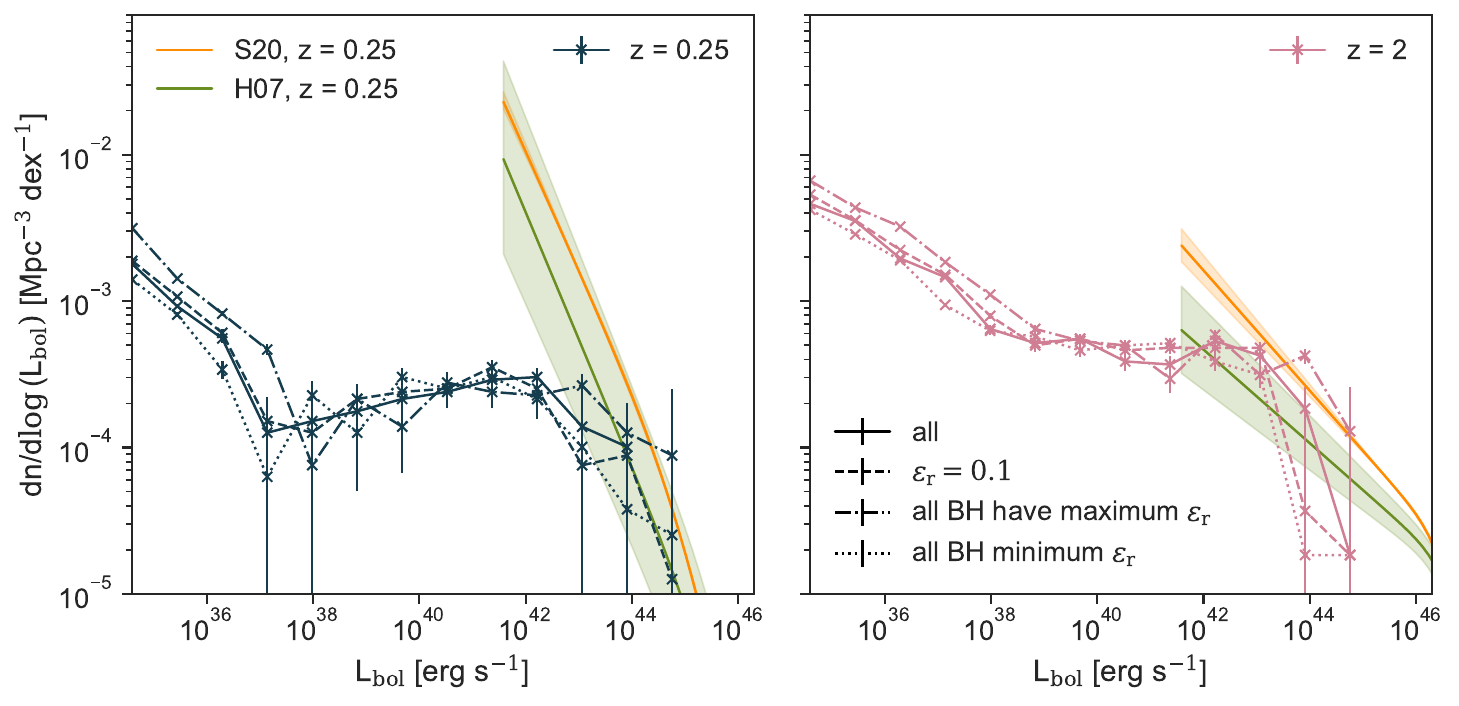}
    \caption{Bolometric BH luminosity function at $z=0.25$ (left) and $z=2$ (right) for all BHs in \nh. The luminosity function is shown for three different assumptions: BHs have the spin-dependent radiative efficiency $\varepsilon_{r}$ as extracted from the simulation (solid), all BHs have a fixed radiative efficiency of $\varepsilon_{r} =0.1$ (dashed), all BHs have a spin-dependent $\varepsilon_{r} =0.1$ and are assumed to be maximally (dash-dotted) and non-spinning (dotted) respectively. Observations from \citep{soares&nemmen20} (S20) and \citep{Hopkins2007} (H07) are shown for comparison.}
    \label{fig:luminosity}
\end{figure*}

In Fig. \ref{fig:luminosity} we test whether using a spin-based $\varepsilon_{\rm r}$ has a significant impact on the bolometric AGN luminosity function by comparing the spin-based luminosity function (solid line) to the one that assumes $\varepsilon_{\rm r}=0.1$ for all BHs, as in previous comparable work. As can be seen by comparing the two lines, the difference in bolometric luminosity is negligible, both at high ($z=2$) and low ($z=0.25$) redshift. The fact that \nh~generally under-predicts the luminosity function at low redshift is explored further in \citet{Beckmann2023}. Two further lines in Fig. \ref{fig:luminosity} show that even under extreme assumptions (all BHs are maximally radiatively efficient, dash-dotted line and all BHs are minimally radiatively efficient, dotted line, assuming their current accretion disc structure) the impact of a spin-dependent $\varepsilon_{\rm r}$ on the luminosity function remains small (within a factor of 3) for the sample of BHs in \nh. 

We caution that the sample of BHs in \nh~is dominated by intermediate-mass, low-luminosity AGN. Conclusions on the impact of a spin-dependent radiative efficiency on the luminosity function might change significantly for a more representative sample. The consequences of increasing the radiative efficiency for massive BHs might require re-calibration of models elsewhere. For example, \citet{Sijacki2015x} found that using an average radiative efficiency of $\varepsilon_r = 0.2$ over-predicted the AGN luminosity function, with $\varepsilon_r=0.1$ in better agreement with observations (but also readily acknowledge that in their model $\varepsilon_r$ is degenerate with a numerical feedback efficiency similar to $\varepsilon_f$). Overall, a careful study of a large sample of massive BHs with on-the-fly spin evolution will be needed to evaluate the impact of spin-dependent radiative efficiency on BH-galaxy coevolution.

\section{Black hole spin observability}
\label{sec:observability}

To be observable, BHs have to be sufficiently luminous to meet observational luminosity cuts. In this section, we explore how the observable distribution of BH spins can vary over time due to a combination of evolving BH luminosity and BH accretion state. As discussed briefly in  Section \ref{sec:spin}, different observational methods probe BHs in different accretion states.

Looking at the final redshift of the simulation ($z=0.25$), the potentially observable sample is small: out of a sample of 583 BHs, only 6 exceed a bolometric luminosity of $10^{39} \rm \, erg \, s^{-1}$, with none brighter than $3 \times 10^{43} \rm \, erg \, s^{-1}$. This is because our sample of BHs is dominated by intermediate-mass BHs (see Fig. \ref{fig:spin_merger_mass} and \citealp{Beckmann2023} for a full analysis) and the sample of massive BHs is small. All 6 bright BHs are in jet mode, which is not unexpected as there are only four BHs in quasar mode within the whole sample at this point (see Fig. \ref{fig:spin_merger_mass}).

\begin{figure*}
    \centering
    \includegraphics[width=\textwidth]{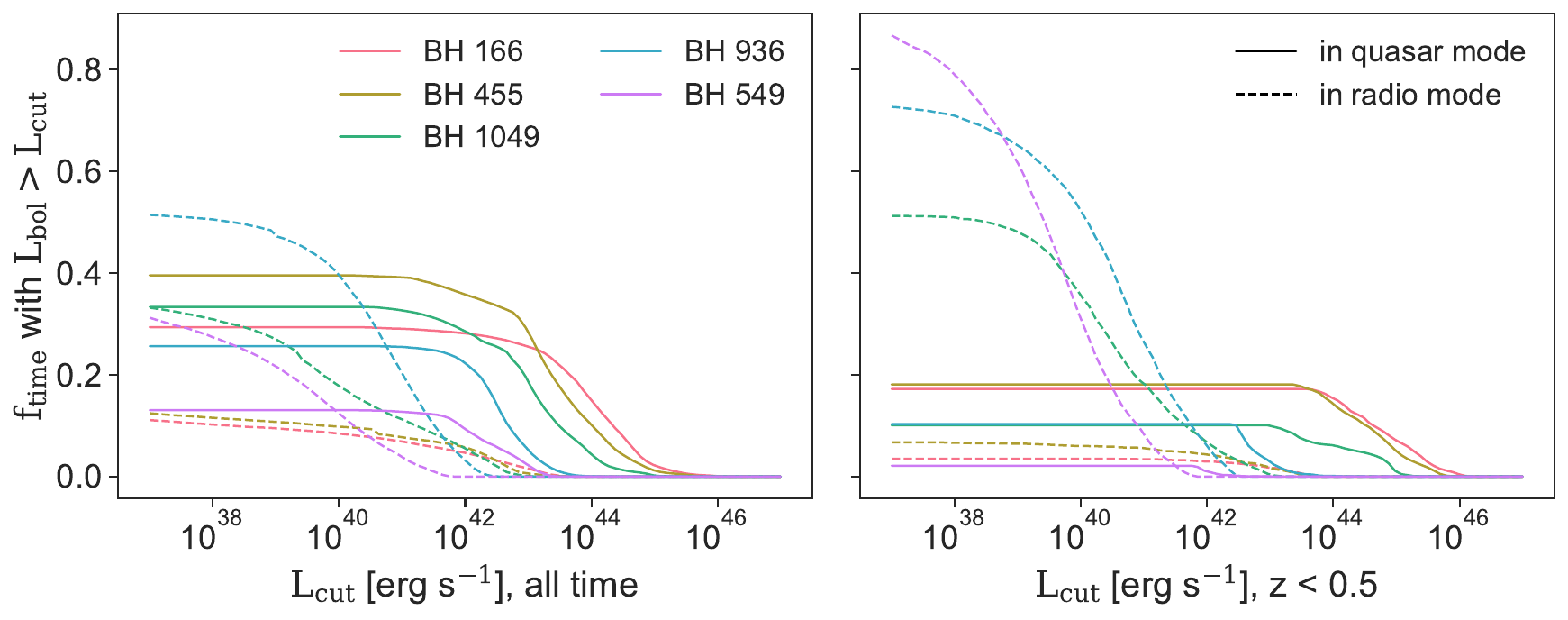}
    \caption{Fraction of time $\rm f_{\rm time}$ individual BHs spend in quasar (solid) or radio (dashed) mode with a bolometric luminosity above a given luminosity cut. Fractions are measured across cosmic history (left) and in the range $z=0.5$ to $z=0.25$ (right).}
    \label{fig:accretion_percentage}
\end{figure*}

However, BHs show strong variability over time, so when exactly their luminosity and accretion state is measured can make a significant difference to their observability, as can be seen in the example timeseries in Sec. \ref{sec:spin}. We quantify this variability for the five most massive BHs in \nh~in Fig. \ref{fig:accretion_percentage}, which shows the percentage of time a BH would be observable in either quasar mode or radio mode for a given luminosity cut. Fractions are measured for all time (left) and since $z<0.5$ only (right). At high luminosity, the probability of observing any of the BHs in \nh~drops significantly, but all observable BHs will be in quasar mode, so only observable with X-ray-based methods. At lower luminosity ($< 10^{42} \, \rm erg \, s^{-1}$), some BHs are much more likely to be in a jetted radio-mode, especially at $z<0.5$ (e.g. BH 936 and 549) while others have a constant probability of being observed in quasar mode (here around 20 per cent for BHs 166 and 455) but a small probability of being found in radio mode at any reasonable luminosity cut. 

It has recently been reported that all massive galaxies with stellar masses $M_{\rm star} > 10^{11} \, M_\odot$ in the local Universe show radio emission associated with the AGN \citep{sabaterLoTSSViewRadio2019}. At first glance, we struggle to reproduce this result as our most massive BHs (BHs 166 and 455), which are also hosted in our most massive galaxies, spend more time in quasar mode than radio mode at low redshift. At any given time, the fraction of AGN in radio mode in galaxies with stellar masses $> 10^{11} M_\odot$ in \nh~is therefore less than 100\%. This could be a sign that BH growth in \nh~is artificially delayed or reduced, as was also suggested by the fact that all massive BHs in \nh~lie in the lower region of the $M_{\rm bh} - \sigma$ relation, and that \nh~generally fails to reproduce the observed fraction of AGN in dwarf galaxies the local Universe \citep[see][for more information]{Beckmann2023a}. It could also mean that our binary disc model is too abrupt, and a smoother transition between quasar mode and radio mode via a truncated disc, such as in \citet{koudmaniUnifiedAccretionDisc2024} is required, or that radio emitting features driven by AGN are more long-lived than the short-scale variability in their accretion mode. Given that our sample of such massive galaxies, and their BHs is extremely small, we note the importance of using such population-wide observations to constrain BH accretion but post-pone a careful study of its impact to future work with a sample better suited to do population-wide studies.

\begin{figure*}
    \centering
    \begin{tabular}{cc}
        \includegraphics[width=0.45\textwidth]{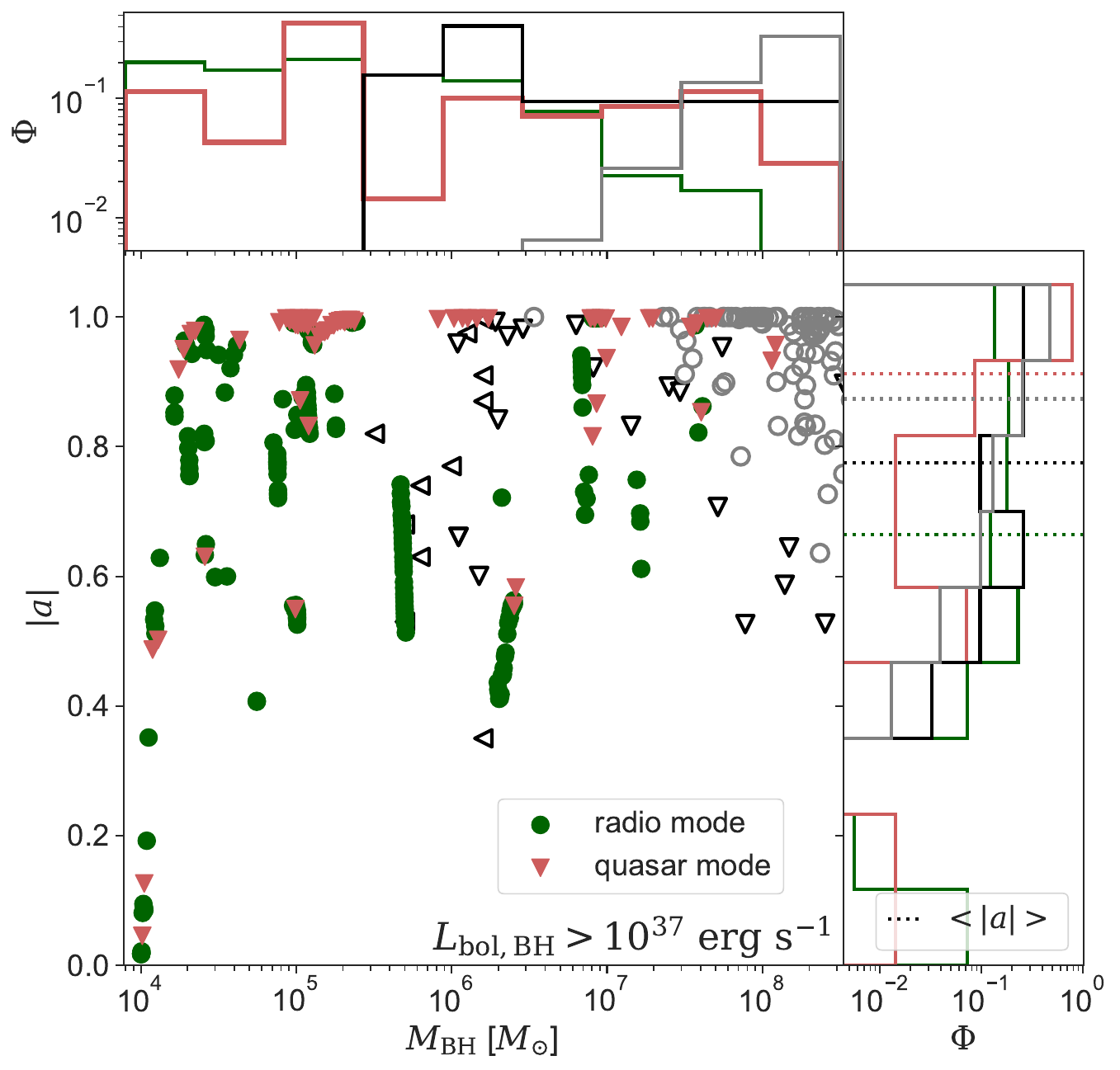} &
         \includegraphics[width=0.45\textwidth]{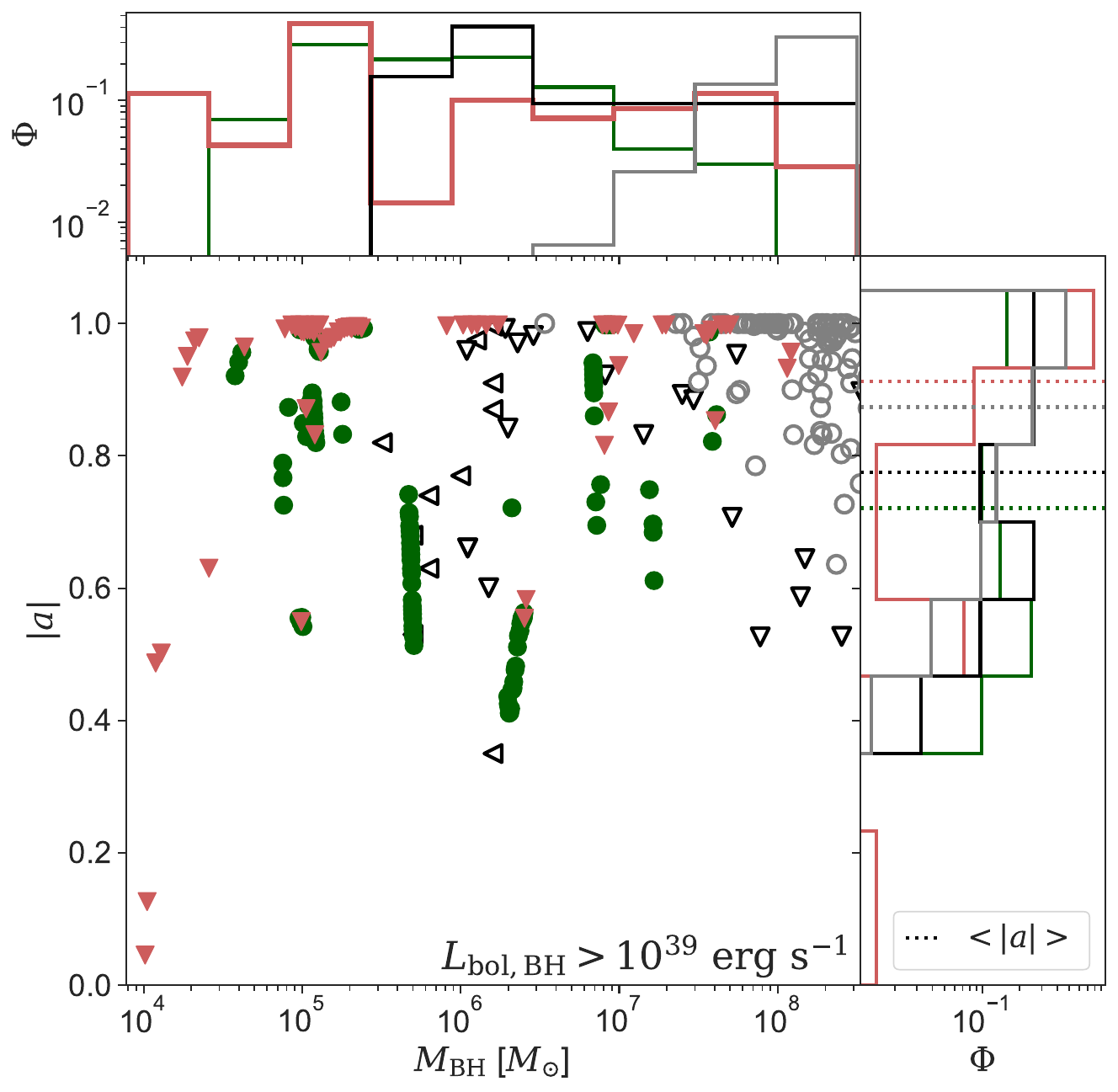}\\
          \includegraphics[width=0.45\textwidth]{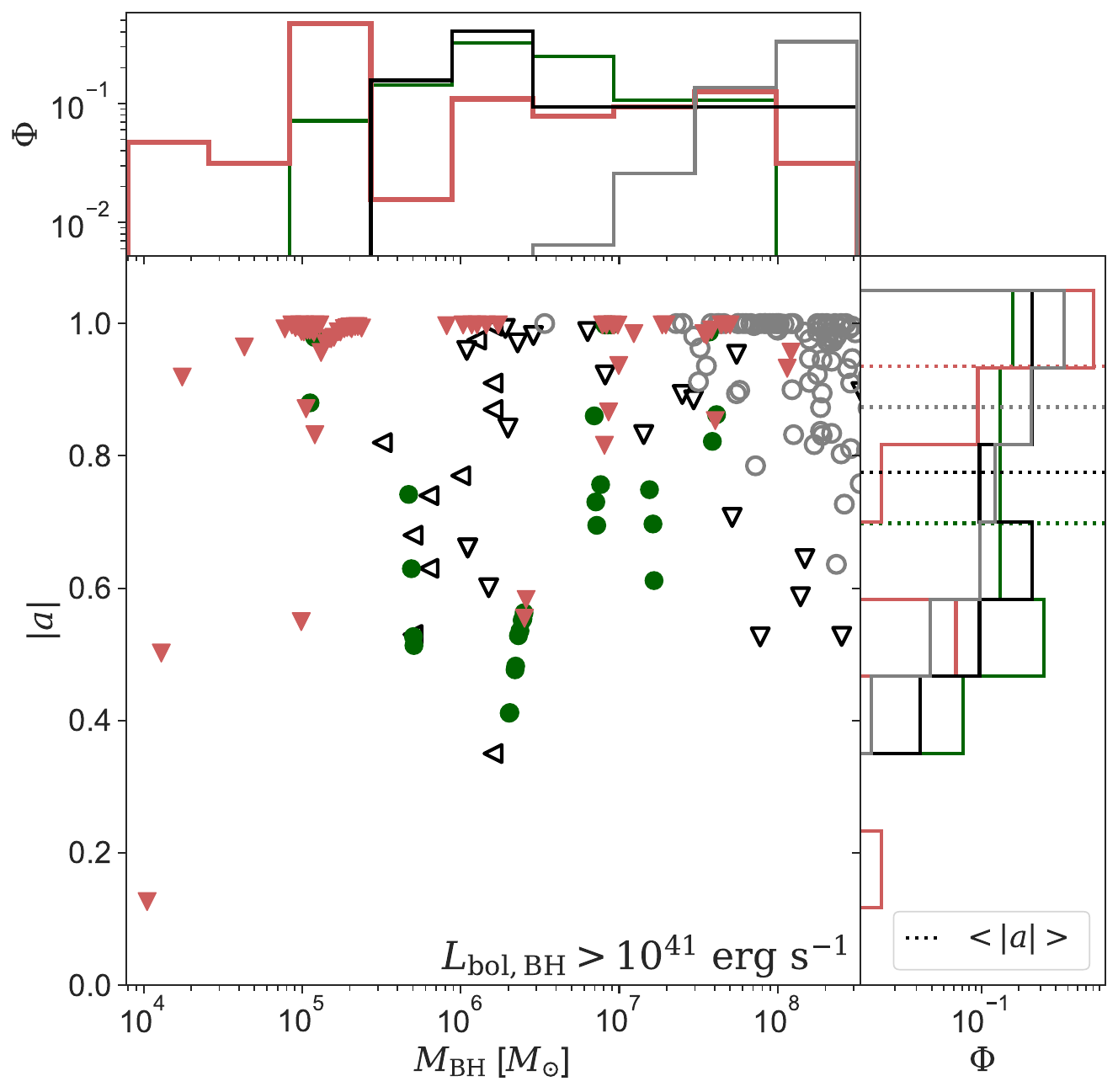} &
          \includegraphics[width=0.45\textwidth]{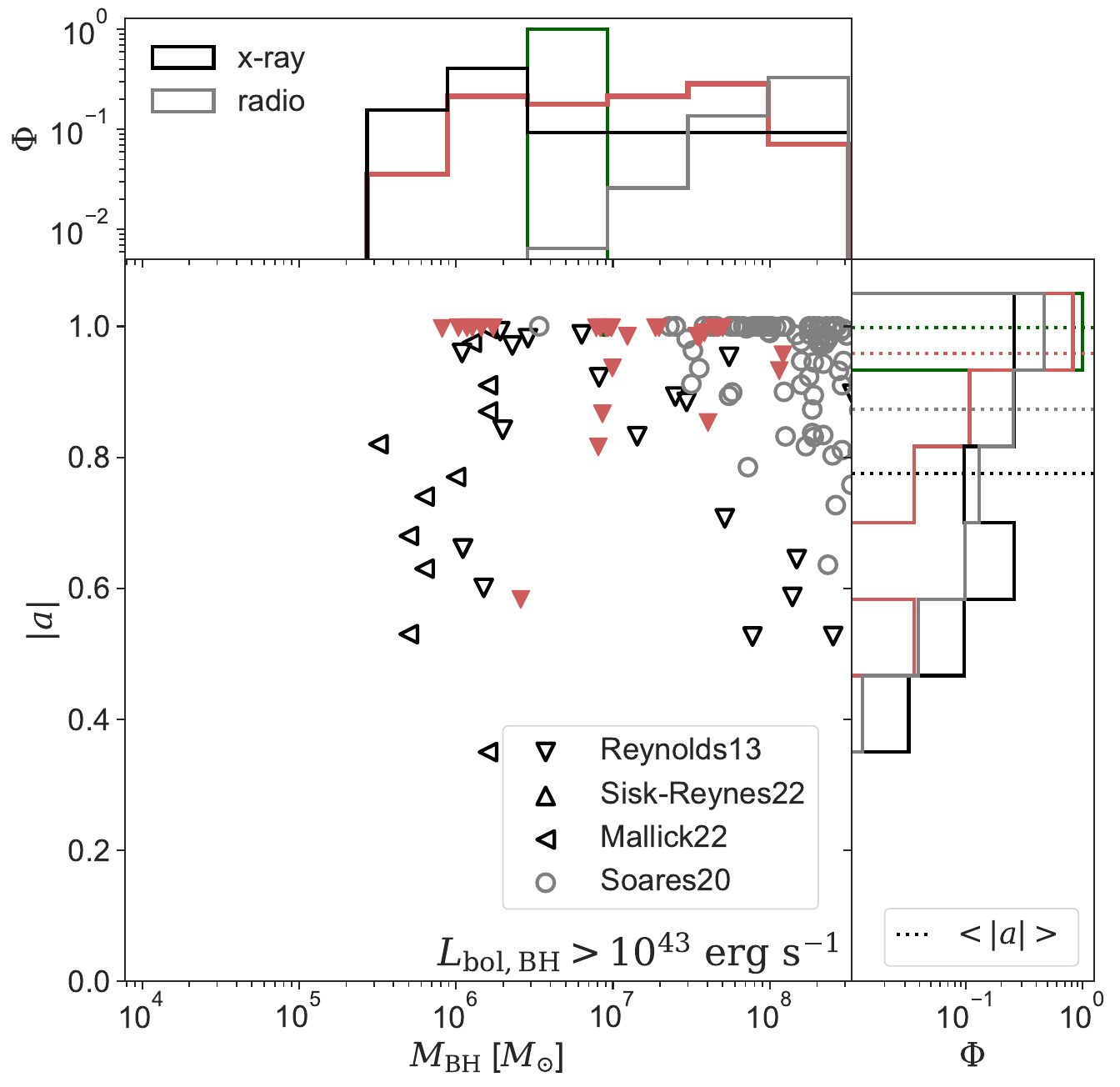} \\

    \end{tabular}
    \caption{Distribution of observable spins from both quasar mode (red triangles) and jet mode (green circles) for different luminosity cuts. All \nh~ BHs are sampled 30 times in the redshift range $z=0.5$ to $z=0.25$ to account for variability in both luminosity and accretion state. Shown for comparison are observed BH spins in the local Universe for X-ray-based (empty triangles, \citep{reynolds_observational_2020,sisk-reynes_evidence_2022,mallick_high-density_2022}) and radio-based techniques (empty circles, \citep{soares&nemmen20}). No luminosity cut has been applied to observations beyond that of the original observational sample.}
    \label{fig:observability}
\end{figure*}

Shown in Fig. \ref{fig:observability} is the population of observable BHs in both radio and quasar mode for a stacked sample of BHs with different minimum bolometric luminosity cuts. All BHs in \nh~have been sampled 30 times at even time intervals from $z=0.5$ to $z=0.25$ to account for variability in both accretion state and luminosity. As can be seen in Fig. \ref{fig:observability}, not surprisingly the distribution of observable BHs shifts towards higher masses with increasing minimum luminosity. At the same time, higher luminosity cuts mean a shift towards BHs in quasar mode, and a decreasing sample size for BHs observable via their radio jets. It also shifts the observed BH spins to higher spin values, with low-spin BHs dropping out of the observable sample. This includes both low-mass BHs with masses $M_{\rm BH} < 10^5\, \rm M_\odot$, which can have spins as low as $|a|<0.2$, and higher mass BHs with masses $M_{\rm BH} > 10^6 \, \rm M_\odot$ in radio mode that tend to have spins in the range $|a| = 0.4 - 0.8$. In comparison to observations, we find fewer quasar-mode BHs in this mass and spin range than reported by X-ray-based observations, but we caution that our sample is too small to draw firm conclusions.

Overall we conclude that the required high luminosity cuts (e.g. \citet{reynolds13}  only consider sources with a minimum of 0.01 the Eddington luminosity, equivalent to $L > 1.3 \times 10^{42} \rm \ erg \ s^{-1}$ for a $10^6 \rm \ M_\odot$ BH) for observations of BH spin bias the distribution of spins towards higher spin values than the full sample of BHs \citep[see also a similar discussion in ][]{Brenneman2011}. 
We caution that the sample of BHs in \nh~is too small for statistical analysis, and the numbers quoted are highly specific to our individual BHs. This discussion is therefore only intended to highlight the variety of BH evolution histories and their potential impact on the observability of BH spin. Further work is required to understand how the observable distribution of BH spin reflects the underlying BH spin distribution for different observational methods, luminosity cuts and galaxy types surveyed \citep[see e.g.][]{Beckmann2023a}. We leave a robust statistical analysis to future work based on a larger-volume sample.

\section{Conclusions}
\label{sec:conclusions}

In this paper, we studied the spin evolution of a sample of BHs in the mass range $10^4 - 10^8 \, \rm M_\odot$ from their formation to $z=0.25$ using the cosmological simulation \nh. We tracked how BH spin changes through gas accretion, BH-BH merger and BH feedback, and concluded that 
\begin{enumerate}
\item BH spin evolution goes through three phases: an early gas-driven spin-up to maximum spin following formation, a merger-induced scattering phase up to a BH mass of $10 \times  \rm M_{BH,0}$, where $M_{\rm BH,0}$ is the BH seed mass (here $M_{\rm BH,0}=10^4 \rm M_\odot$). This is followed by a second accretion-driven phase for BHs in the mass range $10^5 - 10^ 8 \rm M_\odot$. 
\item During the second accretion-driven phase, when BHs are in the mass range of $10^5 - 10^8 \rm M_\odot$, they can be both spun up (due to efficient accretion through a thin disc) or spun down (through inefficient accretion through a thick disc, where BH spin energy is extracted to drive BH jets). Due to the high variability of the accretion rate, all BHs alternate between both modes, significantly increasing the scatter in the spin distribution in this mass range where BHs had been reported as predominantly maximally spinning when jet spindown was not taken into account \citep{Dubois2014spin,Bustamante2019}. 
\item BH-BH mergers for BHs close to their seed mass (here $10^4 \rm \, M_\odot$) predominantly spin-up BHs, while spin typically decreases during mergers between more massive BHs, in agreement with work by \citet{DongPaez2023}. As the BH mass increases the average merger ratio decreases, making BH mergers less important to the long-term spin evolution of the BH. We cannot comment on the previously reported second merger-driven spin-down phase for BHs with masses $M_{\rm BH} > 10^8 \, \rm M_\odot$ (see \citealp{Dubois2014spin} and \citealp{Bustamante2019}) as our BHs are insufficiently massive by the end of the simulation.
\item After an early accretion burst after formation, gas accretion onto BHs undergoes a highly chaotic, inefficient phase during which BH spin (and mass) evolves predominantly through BH-BH mergers. At late times, BH spin is typically well-aligned with the angular momentum of the accreted gas, allowing for significant accretion-driven spin up (or down, depending on the accretion mode and alignment) over Gyr time periods.
\item We find no evidence for spin-down through efficient chaotic accretion onto BHs, either in the early Universe or today. In the early Universe, chaotic accretion is either too inefficient to allow for significant BH growth, or too coherent to reduce BH spin. At late times, we find little evidence for persistent chaotic accretion onto massive BHs at all. We do however report on one episode of persistent anti-aligned accretion over $\sim 2 \,\rm Gyr$ that reduces BH spin.
\item At $<\varepsilon_{\rm r}^{\rm thin}> = 0.19$, the average radiative thin disc efficiency for BHs with masses $M_{\rm BH} > 10^5 \, \rm M_\odot$ is on average almost twice as high as the commonly used value of $0.1$. For BHs in radio mode, even high spin is unable to compensate for the fact that their accretion discs are inherently radiatively inefficient, making such BHs hard to observe.
\item Due to their high spin, massive BHs deposit on average between $\sim 3$ times (in quasar mode) and $\sim 8$ times (in radio mode) more feedback energy into their host galaxy than previously assumed in fixed-spin models.
\item The impact of the spin-based radiative efficiency on the luminosity function is negligible for the sample of BHs presented here.
\item X-ray-based and radio-based observations of BH spin probe  BHs in different feedback modes, with X-ray-based methods probing on average a more highly spinning population of BHs than radio-based approaches.
\item When observing BH spins, higher luminosity cuts lead to higher average observed BH spin values. 
\end{enumerate}

Overall, we conclude that BH spin evolution changes significantly over cosmic time, with both accretion and mergers dominating during different phases of the BH's evolution. This has important consequences for their radiative efficiency, and therefore the feedback energy experienced by the BH's host galaxy. Massive galaxies, in particular, have the potential to experience feedback up to twice as strong as canonically modelled if their SMBH are in quasar mode. As has been explored here, the distribution of observed BH spins depends significantly on the observational method. As the sample studied here is dominated by intermediate-mass BHs, with only a small number of massive BHs up to $\sim 10^8 \, \rm M_\odot$, we leave a robust exploration of this effect to future work.

\section*{Data availability}
Data is available upon request to the corresponding author.

\section*{Acknowledgements}
For the purpose of open access, the author has applied a Creative Commons Attribution (CC BY) licence to any Author Accepted Manuscript version arising from this submission.

This work was granted access to the HPC resources of CINES under the allocations c2016047637, A0020407637 and A0070402192 by Genci, KSC-2017-G2-0003 by KISTI, and as a “Grand Challenge” project granted by GENCI on the AMD Rome extension of the Joliot Curie supercomputer at TGCC.  This research is part of the Spin(e) ANR-13-BS05-0005 (http://cosmicorigin.org), Segal ANR-19-CE31-0017 (http://secular-evolution.org) and Horizon- UK projects. This work has made use of the Infinity cluster on which the simulation was post-processed, hosted by the Institut d’Astrophysique de Paris. We warmly thank S. Rouberol for running it smoothly. The large data transfer was supported by KREONET which is managed and operated by KISTI. RSB would like to thank Newnham College, Cambridge, for financial support. S.K.Y. acknowledges support from the Korean National Research Foundation (NRF-2020R1A2C3003769, NRF-2022R1A6A1A03053472).




\bibliographystyle{mnras}
\bibliography{NewH_spin.bib} 





\bsp	
\label{lastpage}
\end{document}